\documentclass[twocolumn,aps,prb,notitlepage]{revtex4-2}
\pdfoutput=1
\usepackage{bm}
\usepackage{graphicx}
\usepackage{amssymb}
\usepackage{amsmath}
\usepackage{dcolumn}
\usepackage{amsfonts}
\usepackage{upgreek}
\usepackage{comment}
\usepackage[margin=.7in]{geometry}
\usepackage[euler]{textgreek}
\usepackage{lineno}
\usepackage{braket}
\usepackage[caption=false]{subfig}
\usepackage{floatrow}
\usepackage{gensymb}
\usepackage{multirow}
\usepackage{wasysym}
\usepackage{mathtools}
\usepackage[usenames,dvipsnames]{xcolor}
\usepackage[normalem]{ulem}
\usepackage[colorlinks]{hyperref}
\hypersetup{
    colorlinks=true,
    citecolor=blue,
    linkcolor=blue,
    filecolor=magenta,      
    urlcolor=blue,
    }
\newcommand{\stkout}[1]{\ifmmode\text{\sout{\ensuremath{#1}}}\else\sout{#1}\fi}
\newcommand{\bs}{\boldsymbol}
\newcommand{\pd}{\partial}

\begin{document}

\title{Proximity-induced nonlinear magnetoresistances on topological insulators}
\author{M. Mehraeen}
\email{mxm1289@case.edu}
\author{Steven S.-L. Zhang}
\email{shulei.zhang@case.edu}
\affiliation{Department of Physics, Case Western Reserve University, Cleveland, Ohio 44106, USA}
\date{\today}
\begin{abstract}
We employ quadratic-response Kubo formulas to investigate the nonlinear magnetotransport in bilayers composed of a topological insulator and a magnetic insulator, and predict both unidirectional magnetoresistance and nonlinear planar Hall effects driven by interfacial disorder and spin-orbit scattering. These effects exhibit strong dependencies on the Fermi energy relative to the strength of the exchange interaction between the spins of Dirac electrons and the interfacial magnetization. In particular, as the Fermi energy becomes comparable to the exchange energy, the nonlinear magnetotransport coefficients can be greatly amplified and their dependencies on the magnetization orientation deviate significantly from conventional sinusoidal behavior. These findings may not only deepen our understanding of the origin of nonlinear magnetotransport in magnetic topological systems but also open new pathways to probe the Fermi and exchange energies via transport measurements.
\end{abstract}

\maketitle

\section{Introduction}

Topological-insulator (TI)-based magnetic heterostructures are appealing systems for exploring the interplay between magnetism and band topology. These hybrid systems are characterized by the coexistence of strong spin-orbit coupling (SOC), sizable magnetic exchange interaction, and Dirac surface states with spin-momentum locking. Moreover, it has been demonstrated that the Fermi level of the TI layer in these systems can be finely tuned with respect to the Dirac point~\cite{Okada16PRB_Fermi-level_TI, Kondou16NP_Fermi-level-sc_TI, sun2019large, Su21ACS_Fermi-level_TI}. These properties are remarkable in their own rights, and a blend of them makes these systems even more intriguing. A multitude of linear-response transport phenomena have attracted considerable attention, including quantum anomalous Hall~\cite{qi2006topological, yu2010quantized, chang2013experimental, checkelsky2014trajectory, liu2016quantum, chang2023colloquium}, topological Hall~\cite{wu2020ferrimagnetic, li2021topological, zhang2021giant}, spin-transfer torque~\cite{Mellnik2014,Han17PRL_SOT-TI,li2019magnetization}, and various novel magnetoresistance effects~\cite{Moghaddam20PRL_SOT-AMR-proximity-TI,chiba2017magnetic,sklenar2021proximity}, which may potentially lead to applications in many areas, ranging from classical information storage and processing~\cite{Wu21NC_MRAM-TI} to quantum computation~\cite{Fu08PRL_proximity-SC_TI_majorana,Lian18PNAS_Topo-quant-comput_MZM}. 

Going beyond linear responses, TI-based magnetic heterostructures also allow magnetotransport that violates Onsager’s reciprocity in principle, owing to the lack of both time-reversal and inversion symmetries. Corrections to linear magnetoconductivities have been observed in a few TI-based magnetic heterostructures~\cite{yasuda2016large, yasuda2017current, lv2018unidirectional, duy2019giant, wang2022observation, lv2022large}–ensuing the discovery of unidirectional magnetoresistance (UMR) effects in metallic and semiconducting magnetic bilayers~\cite{avci2015unidirectional, olejnik2015electrical}. Such nonlinear magnetoconductivities are odd under the reversal of either the direction of the applied electric field $\mathbf{E}$ or that of the magnetization (whose direction will be denoted by a unit vector $\mathbf{m}$ hereafter); i.e., 
$\sigma^{(2)}(-\mathbf{m},\mathbf{E})
=
\sigma^{(2)}(\mathbf{m},-\mathbf{E})
=
-\sigma^{(2)}(\mathbf{m},\mathbf{E})$, which are distinctly different from their linear-response counterparts and hence hold fascinating prospects for adding new functionalities in future spintronic devices.

To date, studies of nonlinear transport in magnetic layered structures have mainly been focused on controlling the corresponding magnetotransport coefficients by varying the magnetization vector with an external magnetic field. And, more specifically, the reported dependencies of the nonlinear current on the magnetization direction can be cast into the simple form~\cite{olejnik2015electrical, avci2015unidirectional, zhang2016theory, guillet2021large, zelezny2021unidirectional, zhou2021sign, liu2021magnonic, nguyen2021unidirectional, hasegawa2021enhanced, mehraeen2022spin, shim2022unidirectional, lv2022large, lou2022large, ding2022unidirectional, fan2022observation, cheng2023unidirectional, wang2023controllable, zheng2023coexistence, mehraeen2023quantum}
\begin{equation}
\label{j_1^2}
\mathbf{j}_{1}^{(2)}
=
\sigma_{\parallel,1}^{(2)} \mathbf{m} \cdot (\mathbf{z} \times \mathbf{E}) \mathbf{E}
+
\sigma_{\perp,1}^{(2)} (\mathbf{m} \cdot \mathbf{E}) \mathbf{z} \times \mathbf{E},
\end{equation}
where the unit vector $\mathbf{z}$ denotes the interface normal and $\sigma_{\parallel,1}^{(2)}$ ($\sigma_{\perp,1}^{(2)}$) is a transport coefficient that characterizes the strength of the nonlinear current longitudinal (transverse) to the applied electric field, which is denoted by $\mathbf{j}_{\parallel,1}^{(2)}$ ($\mathbf{j}_{\perp,1}^{(2)}$). Here, the superscript 2 and the subscript 1 indicate that the nonlinear current is of second order in the applied electric field and first order in the magnetization. The tunability of these nonlinear transport coefficients upon the shift of the Fermi level, however, has remained unexplored. 

\begin{figure*}[hpt]
\captionsetup[subfigure]{labelformat=empty}
    \sidesubfloat[]{\includegraphics[width=0.32\linewidth,trim={1.7cm 1cm 0.5cm 0.7cm}]{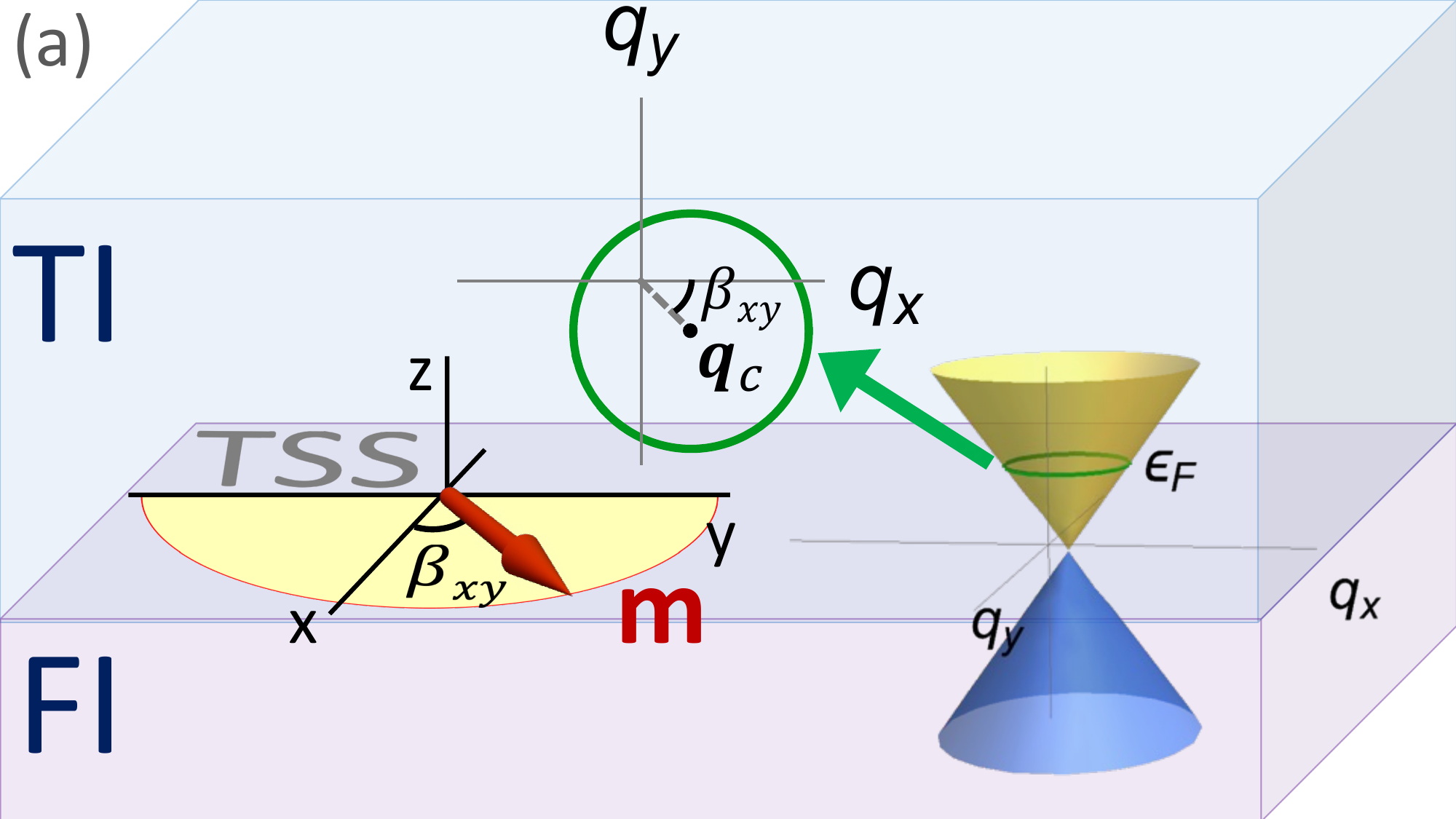}\label{fig1a}}
    \sidesubfloat[]{\includegraphics[width=0.32\linewidth,trim={1.8cm 1cm 0.3cm 0.7cm}]{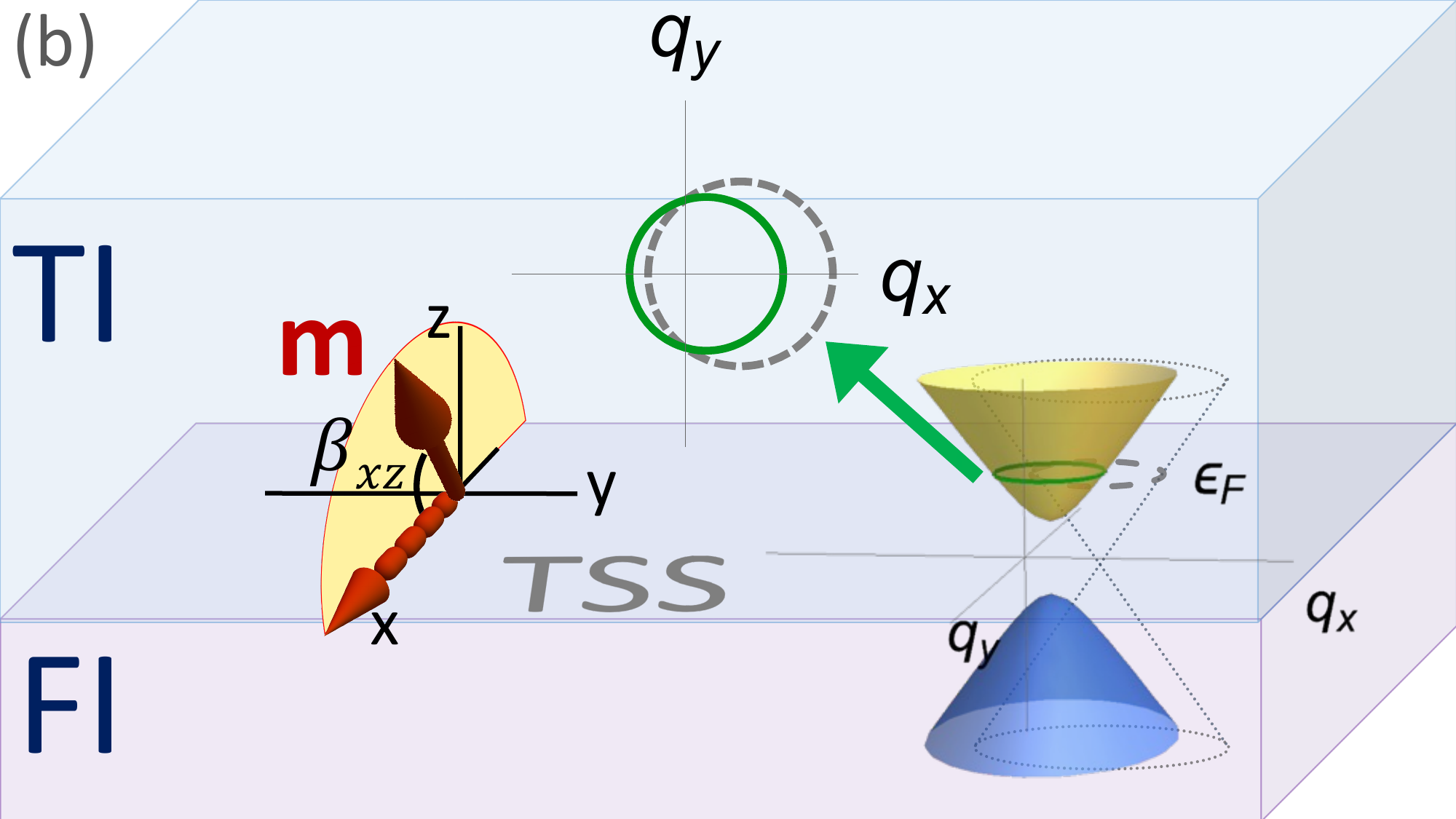}\label{fig1b}}
    \sidesubfloat[]{\includegraphics[width=0.32\linewidth,trim={1.8cm 1cm 0.5cm 0.7cm}]{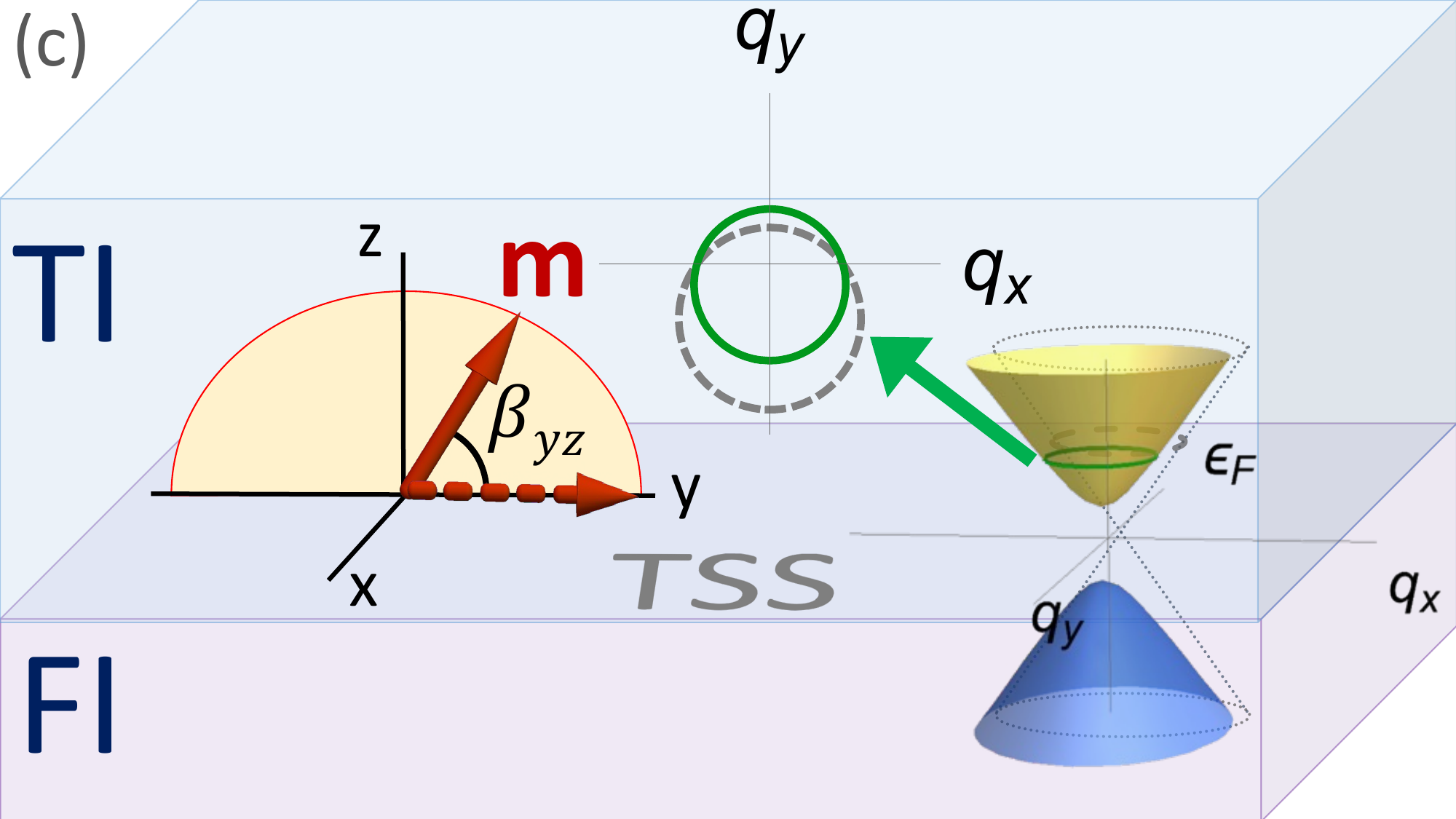}\label{fig1c}}
    \caption{Schematics of the TI/FI bilayer system and the topological surface state (TSS). (a) In-plane rotation of the magnetization, which leads to a rigid shift of the Dirac cone and thus the Fermi contour. (b) and (c) Out-of-plane rotations of the magnetization, which gap the system. In this case, in addition to a shift, the radius of the Fermi contour is decreased as the magnetization is rotated out of the bilayer plane. Here, $\beta_{ij}$ ($i,j=x,y,z$) is the angle between the $i$ axis and the magnetization as it sweeps the $ij$ plane.}
    \label{fig1}
\end{figure*}

In this work, we theoretically investigate nonlinear magnetotransport in magnetic bilayers consisting of a TI and a ferromagnetic-insulator (FI) layer, as shown schematically in Fig.~\ref{fig1}. A formal evaluation of Kubo formulas in the quadratic-response regime is performed to explore the magnetoconductivities of interest. Within this quantum approach, we find a UMR effect and a nonlinear planar Hall effect (NPHE) that are driven purely by extrinsic spin-orbit scattering at the interface. More interestingly, perhaps, unconventional dependencies of the nonlinear currents on the magnetization orientation emerge, which differ from those given by Eq.~(\ref{j_1^2}) and feature terms that are of higher order in the magnetization. Following a symmetry analysis of the nonlinear response function, the details of which are presented in Appendix \ref{appendixA}, the general nonlinear current may be expressed as
\begin{subequations}
\label{j^2_cubic}
\begin{align}
\mathbf{j}_{\parallel}^{(2)}
&\hspace{-.05cm}
=
\hspace{-.03cm}
\mathbf{j}_{\parallel,1}^{(2)}
\left[ 1
\hspace{-.05cm}
+
\hspace{-.05cm}
\iota_{\parallel} (\mathbf{m} \cdot \mathbf{e})^2
\hspace{-.05cm}
+
\hspace{-.05cm}
\kappa_{\parallel} (\mathbf{m} \cdot \mathbf{z} \times \mathbf{e})^2
\hspace{-.05cm}
+
\hspace{-.05cm}
\lambda_{\parallel} (\mathbf{m} \cdot \mathbf{z})^2
\right],
\\
\mathbf{j}_{\perp}^{(2)}
&\hspace{-.05cm}
=
\hspace{-.03cm}
\mathbf{j}_{\perp,1}^{(2)}
\left[ 1
\hspace{-.05cm}
+
\hspace{-.05cm}
\iota_{\perp} (\mathbf{m} \cdot \mathbf{e})^2
\hspace{-.05cm}
+
\hspace{-.05cm}
\kappa_{\perp} (\mathbf{m} \cdot \mathbf{z} \times \mathbf{e})^2
\hspace{-.05cm}
+
\hspace{-.05cm}
\lambda_{\perp} (\mathbf{m} \cdot \mathbf{z})^2
\right],
\end{align}
\end{subequations}
where $\mathbf{e}$ is the unit vector along the electric field. Here, $\iota_{\parallel,\perp}$, $\kappa_{\parallel,\perp}$ and $\lambda_{\parallel,\perp}$ are dimensionless quantities that characterize the strengths of the contributions cubic in the magnetization and are, in general, functions of disorder and the exchange energy. Intriguingly, when the Fermi and exchange energies become comparable, the nonlinear magnetoresistances may be considerably amplified and the contributions of the cubic terms become relatively large, leading to strong deviations in the angular dependencies. These new features may then be used to obtain insights about the position of the Fermi level or the strength of the interfacial exchange interaction via transport measurements, as a simple alternative to optical schemes such as ARPES. 

\section{Disorder scattering in TI surface states}

Let us commence with a minimal model for the surface states on a TI adjacent to a FI layer, which may be expressed as
\begin{subequations}
\label{Ham_main}
\begin{gather}
\label{H_main}
\hat{H}_{\mathbf{q}\mathbf{q}^{\prime}}
=
\hat{H}^0_\mathbf{q} \delta_{\mathbf{q}\mathbf{q}^{\prime}}
+
\hat{V}_{\mathbf{q}\mathbf{q}^{\prime}},
\\
\label{H0_main}
\hat{H}^0_\mathbf{q}
=
\hat{\bs{\sigma}} \cdot \mathbf{h}_{\mathbf{q}}
+
\Delta_{ex} \hat{\sigma}_z m_z,
\end{gather}
\end{subequations}
where $\mathbf{h}_{\mathbf{q}}
=
\hbar v_F \mathbf{q} \times \mathbf{z}
-
\Delta_{ex} \mathbf{z} \times \left( \mathbf{z} \times \mathbf{m}\right)$, with $\mathbf{q}$ the in-plane momentum, $v_F$ the Fermi velocity, $\Delta_{ex}$ the proximity-induced exchange energy and $\mathbf{m} = (m_x, m_y, m_z)$ the unit magnetization. For the impurity potential $\hat{V}_{\mathbf{q}\mathbf{q}^{\prime}}$, we assume it consists of contributions from scalar point scatterers, $\hat{U}_{\mathbf{q}\mathbf{q}^{\prime}}$, as well as from the SOC of the random structural defects \footnote{It should be stressed that the SOC disorder is a required element, as without it, the in-plane magnetization--a key UMR ingredient--can be gauged out of the problem \cite{chiba2017magnetic, dyrdal2020spin}.}, $\hat{W}_{\mathbf{q}\mathbf{q}^{\prime}}$, as \cite{sherman2003minimum, golub2004spin, strom2010edge, kimme2016backscattering, dyrdal2020spin}
\begin{subequations}
\begin{align}
\hat{V}_{\mathbf{q}\mathbf{q}^{\prime}}
&=
\hat{U}_{\mathbf{q}\mathbf{q}^{\prime}}
+
\hat{W}_{\mathbf{q}\mathbf{q}^{\prime}},
\\
\hat{U}_{\mathbf{q}\mathbf{q}^{\prime}}
&=
U_{\mathbf{q}\mathbf{q}^{\prime}}^0 \hat{\sigma}_0,
\\
\hat{W}_{\mathbf{q}\mathbf{q}^{\prime}}
&=
\frac{1}{2} W_{\mathbf{q}\mathbf{q}^{\prime}}^0
\hat{\bs{\sigma}} \cdot \left(\mathbf{q}+\mathbf{q}^{\prime}\right) \times \mathbf{z}.
\end{align}
\end{subequations}
And we assume the white noise distribution for the disorder potentials, $\braket{U_{\mathbf{q}\mathbf{q}^{\prime}}^0}=0$, $\braket{U_{\mathbf{q}\mathbf{q}^{\prime}}^0 U_{\mathbf{q}^{\prime}\mathbf{q}}^0}= n_I U_0^2$, $\braket{W_{\mathbf{q}\mathbf{q}^{\prime}}^0}=0$, and $\braket{W_{\mathbf{q}\mathbf{q}^{\prime}}^0 W_{\mathbf{q}^{\prime}\mathbf{q}}^0}= n_{\alpha} W_0^2$, where $\braket{\cdots}$ denotes the impurity average and $n_I$ and $n_{\alpha}$ are the densities of the scalar and SOC scatterers, while $U_0$ and $W_0$ measure the strengths of the disorder interactions.

\section{Scattering time and quadratic response}

\begin{figure}[tph]
\vspace{.3cm}
{\includegraphics[width=0.6\linewidth,trim={1.5cm 0.5cm 0.5cm 0}]{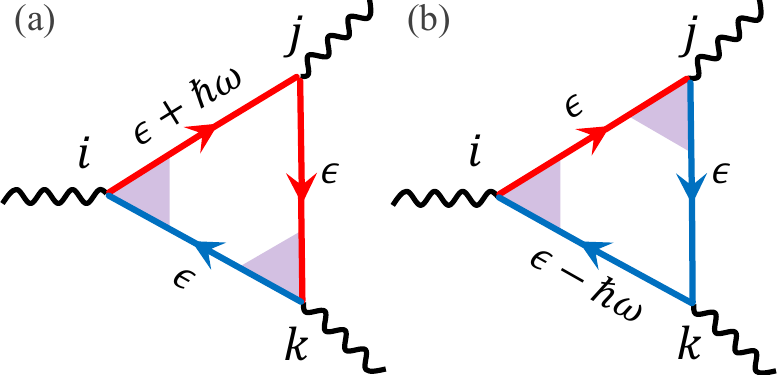}}
	\caption{Diagrammatic structure of the quadratic response. Panels (a) and (b), along with their $j \leftrightarrow k$ counterparts, are the four triangle diagrams that give rise to the UMR and NPHE. The red (blue) arrowed lines represent retarded (advanced) dressed electron Green's functions, while purple shaded areas indicate dressed vertices. The vertices are labeled by the spatial index of their external photons, while propagators are labeled by their energies.}
    \label{fig2}
\end{figure}

In this section, the nonlinear magnetotransport coefficients in the system under consideration are examined by evaluating quadratic Kubo formulas, which--diagrammatically--correspond to \textit{triangle} diagrams of response theory \cite{kubo1957statistical, parker2019diagrammatic, du2021quantum, rostami2021gauge}, as shown in Fig.~\ref{fig2}. This is an essential diagrammatic approach, as UMRs and nonlinear Hall effects cannot be captured by the two-photon bubble diagrams of linear response theory.

The self-energy in the Born approximation is given by
\begin{equation}
\label{selfenergy}
\Sigma^{\text{R/A}}_{\mathbf{q}\sigma}(\epsilon)
=
\sum_{\mathbf{q}^{\prime}\sigma^{\prime}}
\Braket{V_{\mathbf{q}\mathbf{q}^{\prime}}^{\sigma \sigma^{\prime}} V_{\mathbf{q}^{\prime}\mathbf{q}}^{{\sigma^{\prime} \sigma}}}
G^{0,\text{R/A}}_{\mathbf{q}^{\prime}\sigma^{\prime}}(\epsilon),
\end{equation}
where
$\sum_{\mathbf{q}} \equiv \int d^2 \mathbf{q}/(2\pi)^2$ and $G^{0,\text{R/A}}_{\mathbf{q}\sigma}(\epsilon)
=
(\epsilon - \epsilon_{\mathbf{q}\sigma} \pm i \delta)^{-1}$ 
is the retarded/advanced Green's function of the unperturbed system $\hat{H}_{\mathbf{q}}^0$, with the eigenvalue $\epsilon_{\mathbf{q}\sigma}=\sigma \sqrt{h_{\mathbf{q}}^2 + \Delta_{ex}^2 m_z^2}$ for the band $\sigma$ ($=\pm$). The scattering time is given by $\tau_{\mathbf{q}\sigma}(\epsilon)= \hbar/2 \Gamma_{\mathbf{q}\sigma} (\epsilon)$, with the scattering rate defined as $\Gamma_{\mathbf{q}\sigma} \equiv - \text{Im} \Sigma^{\text{R}}_{\mathbf{q}\sigma}$. Upon introducing the change of variables $\bs{\eta}_{\mathbf{h}}\equiv \mathbf{h}_{\mathbf{q}}/\epsilon$ and $\eta_{ex} \equiv \Delta_{ex}/\epsilon$, as detailed in Appendix \ref{appendixB}, we find
\begin{equation}
\label{tau_qsigma}
\tau_{\mathbf{q}\sigma}
=
\frac{\hbar}{2 \Gamma_I}
\left[1+ \sigma \eta_{\text{ex}} m_z \cos \theta_{\mathbf{h}}
+
\frac{\Gamma_{\alpha}}{\Gamma_I} 
\mathcal{F}_{\sigma}
\left( \bs{\eta}_{\mathbf{h}}, \eta_{ex} ; \mathbf{m} \right)\right]^{-1},
\end{equation}
where $\Gamma_I \equiv n_I U_0^2 \epsilon / (2\hbar v_F)^2$ and  $\Gamma_{\alpha} \equiv n_{\alpha} W_0^2 \epsilon^3 /(2\hbar v_F)^4$ are the scalar and SOC disorder self-energy coefficients, respectively, and $\theta_{\mathbf{h}}
=
\cos^{-1}(\eta_{ex} m_z/\sqrt{\eta_{\mathbf{h}}^2 + \eta_{ex}^2 m_z^2})$. The dimensionless band-dependent function $\mathcal{F}_{\sigma}$--whose explicit form is presented in Appendix \ref{appendixB}--is a rather complicated function of the Fermi velocity, exchange energy and orientation of the magnetization. As we discuss below, this nontrivial angular dependence of the scattering time on the magnetization direction plays an important role in explaining the unconventional angular dependencies of the nonlinear magnetoresistances.

The quadratic conductivity tensor is obtained by evaluating the four triangle diagrams shown in Fig.~\ref{fig2}. Together, their contributions to the nonlinear dc conductivity may be succinctly expressed as \cite{du2021quantum}
\begin{widetext}
\begin{equation}
\label{sigma_ijk}
\sigma^{ijk}
=
\frac{e^3 \hbar^2}{\pi}
\text{Im} \sum_{\mathbf{q}\sigma} 
\pd_{\omega}\left[ 
\mathcal{V}^i_{\mathbf{q}\sigma} \left(\epsilon_F, \epsilon_F + \hbar \omega \right)
G^R_{\mathbf{q}\sigma} \left( \epsilon_F + \hbar \omega \right)
\right]_{\omega=0}
v^j_{\mathbf{q}\sigma}
G^R_{\mathbf{q}\sigma} \left( \epsilon_F \right)
\mathcal{V}^{kF}_{\mathbf{q}\sigma}
G^{A}_{\mathbf{q}\sigma} \left( \epsilon_F \right)
+
\left( j \leftrightarrow k\right).
\end{equation}
\end{widetext}
Here $\epsilon_F$ is the Fermi energy and $G^{\text{R/A}}_{\mathbf{q}\sigma}(\epsilon)=(\epsilon - \epsilon_{\mathbf{q}\sigma} \pm i \Gamma_{\mathbf{q}\sigma})^{-1}$ is the disorder-dressed electron Green's function. The (bare) velocity operator is defined as $\hat{\mathbf{v}}_{\mathbf{q}}=\bs{\pd}_{\mathbf{q}}\hat{H}_{\mathbf{q}}^0/\hbar$, with $\pd_{\mathbf{q}}^i \equiv \pd/\pd q_i$, which--in the chiral basis of Bloch eigenstates--leads to the diagonal terms $\mathbf{v}_{\mathbf{q}\sigma}=\pd_{\mathbf{q}}\epsilon_{\mathbf{q}\sigma}/\hbar$. And $\bs{\mathcal{V}}_{\mathbf{q}\sigma} \left(\epsilon, \epsilon^{\prime}\right)$, which is presented in Appendix \ref{appendixC}, is the disorder-averaged velocity vertex function, where $\epsilon$ and $\epsilon^{\prime}$ are, respectively, the energies of the incoming and outgoing propagators to the vertex in question and $\bs{\mathcal{V}}_{\mathbf{q}\sigma}^F \equiv \bs{\mathcal{V}}_{\mathbf{q}\sigma} \left(\epsilon_F, \epsilon_F \right)$.

\section{Angular dependencies}

Without loss of generality, let us set the electric field along the $x$ direction, $\mathbf{e}= \mathbf{x}$. Then it suffices to calculate the $\sigma_{xxx}$ and $\sigma_{yxx}$ elements of the conductivity tensor. To characterize the nonlinear transport, we introduce the longitudinal and transverse UMR coefficients $\zeta_{\parallel}^{(2)} = \zeta_x^{(2)}$ and $\zeta_{\perp}^{(2)} = \zeta_y^{(2)}$, where
\begin{equation}
\label{Eq:zeta}
\zeta_i^{(2)}
\equiv
\frac{\sigma_{ix}(E_{x}) - \sigma_{ix}(-E_{x})}{\sigma_D E_x}
\simeq
-\frac{2 \sigma_{ixx}}{\sigma _{D}},
\end{equation}
to leading order in the electric field. Here $\sigma_{ij}=j_{i}/E_{j}$ denotes the linear conductivity tensor and $\sigma_D= e^2/ [4 \pi \hbar (\eta_I+\eta_{\alpha})]$ is the Drude conductivity, with $\eta_{I} \equiv \Gamma_I/ \epsilon_F$ and $\eta_{\alpha} \equiv \Gamma_{\alpha}/ \epsilon_F$ the dimensionless disorder coefficients.

\begin{figure}[hpt]
    \sidesubfloat[]{\includegraphics[width=0.42\linewidth,trim={2cm 0cm 0.3cm 1cm}]{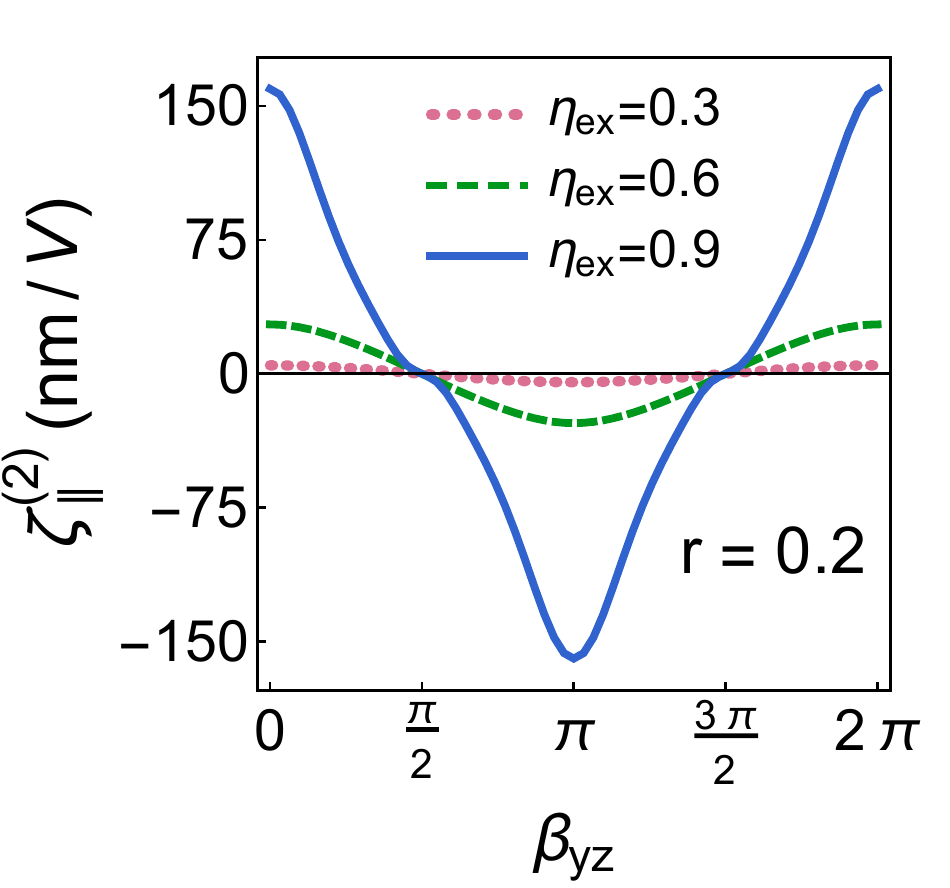}\label{fig3a}}
    \sidesubfloat[]{\includegraphics[width=0.42\linewidth,trim={2cm 0cm 0.6cm 1cm}]{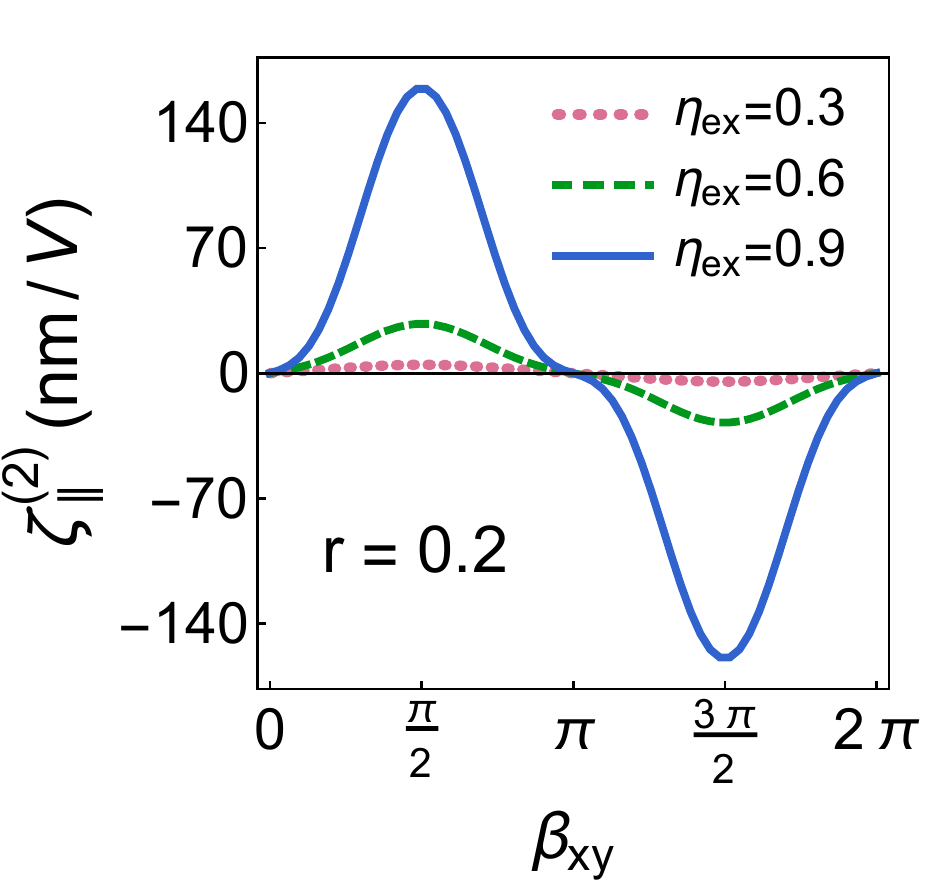}\label{fig3b}}
\\
    \sidesubfloat[]{\includegraphics[width=0.42\linewidth,trim={1.4cm 0.9cm 0.2cm .5cm}]{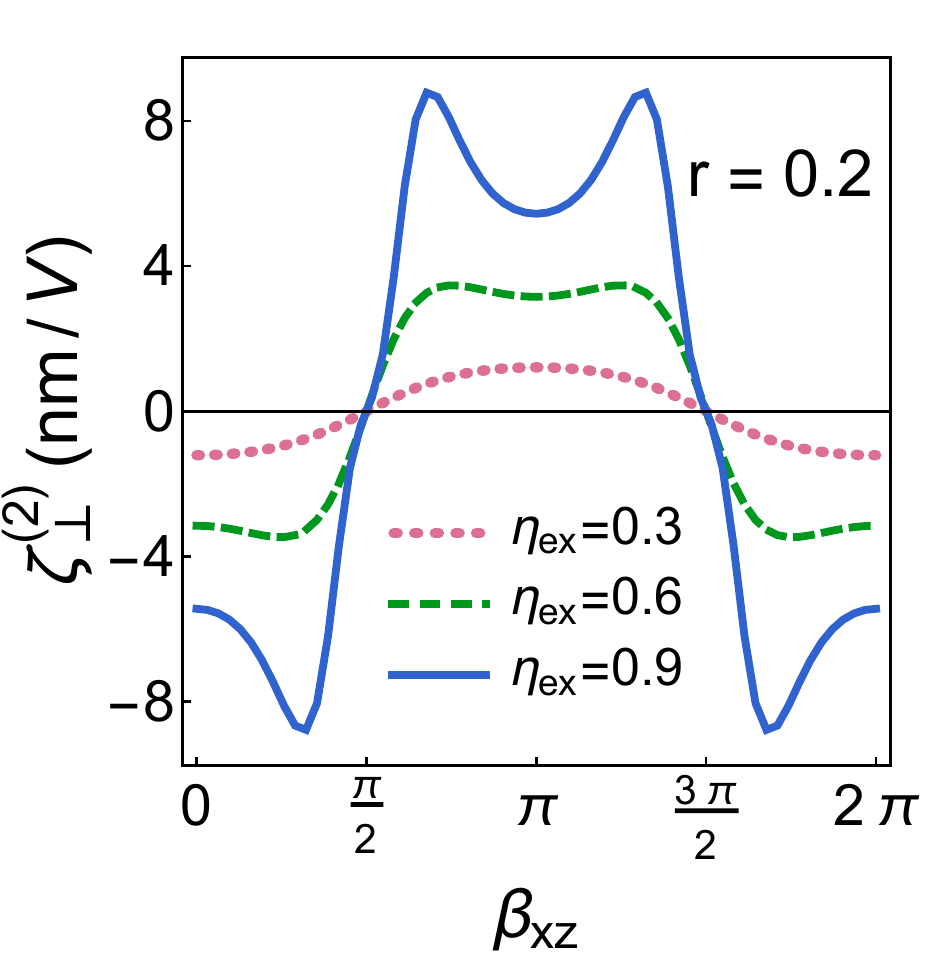}\label{fig3c}}
    \sidesubfloat[]{\includegraphics[width=0.42\linewidth,trim={1.4cm 0.9cm 0.7cm .5cm}]{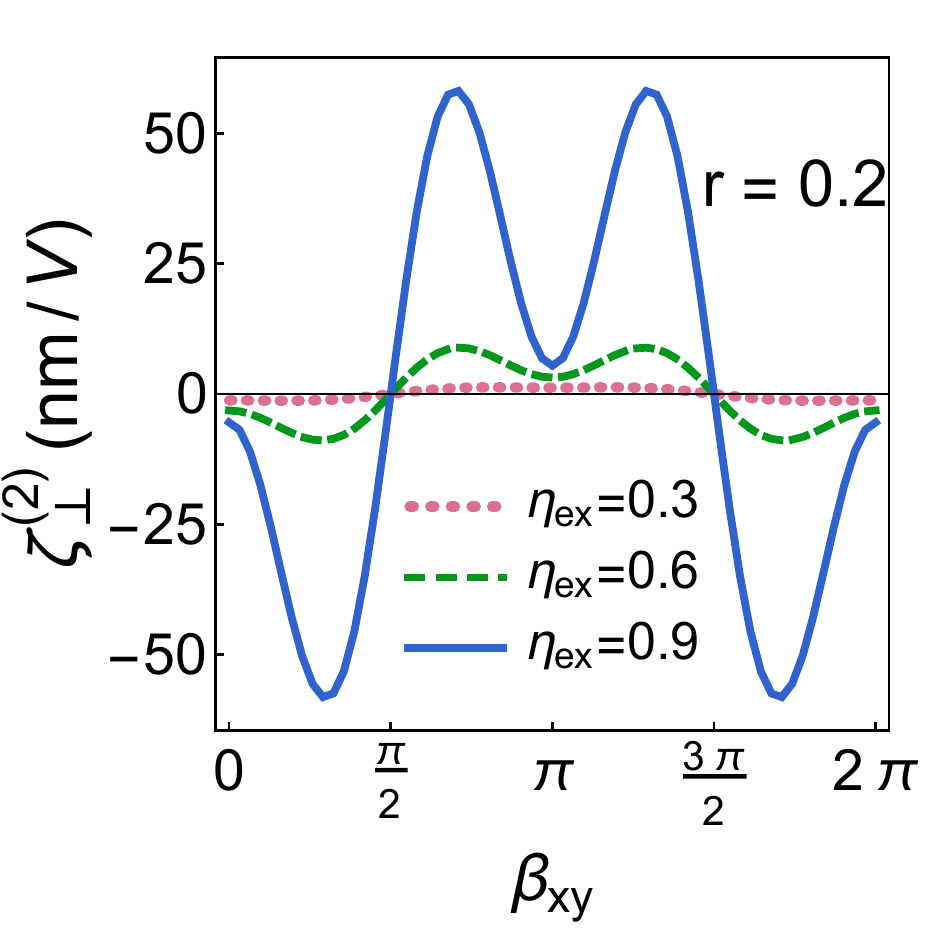}\label{fig3d}}
    \caption{Angular dependencies of the UMR coefficients as the magnetization direction is varied in the $xy$, $yz$, and $xz$ planes (the $xz$ scan for $\zeta_{\parallel}^{(2)}$ and $yz$ scan for $\zeta_{\perp}^{(2)}$ vanish due to symmetry constraints and hence are not shown here). Parameters used: $\epsilon_F=0.5$ eV, $v_F= 5\times10^{14}$ nm/s \cite{qi2011topological} and $\eta_I = 0.01$.}
    \label{fig3}
\end{figure}

Plots of the UMR coefficients for various angular sweeps of the magnetization are presented in Fig.~\ref{fig3}. As shown by the blue curves in Figs.~\ref{fig3a} and \ref{fig3b}, we see that as the Fermi level approaches the exchange energy, \textit{i.e.}, when $\eta_{ex}$ is closer to 1, the strength of the longitudinal nonlinear magnetoresistance $\zeta_{\parallel}^{(2)}$ is significantly amplified and can be as large as 1-2 orders of magnitude stronger than when $\eta_{ex}$ is smaller (see the dashed green and dotted pink curves). Furthermore, qualitatively, the angular dependencies of the longitudinal UMR coefficient increasingly deviate from the conventional sinusoidal behavior as $\eta_{ex}$ is increased.

Quantitatively, the situation is similar for the NPHE; as shown by the blue curves in Figs~\ref{fig3c} and \ref{fig3d}, the transverse UMR coefficient $\zeta_{\perp}^{(2)}$ is larger when the Fermi level is closer to the exchange energy. Qualitatively, however, as $\eta_{ex}$ is increased, the NPHE no longer reaches its maximal absolute value at the expected angles $\beta_{xz}, \beta_{xy}=0,\pi$ (see the pink curves), which is when the magnetization is entirely along the $x$ axis; instead, there is an emergent bifurcation of the peaks and troughs such that the maximal absolute values of the NPHE are obtained when the magnetization is only partially along the $x$ axis \footnote{It is worth noting that these angular dependence trends of the NPHE for both out-of-plane and in-plane sweeps of the magnetization resemble the behavior predicted for the planar Hall effect in the linear response regime \cite{chiba2017magnetic}.}. 

Further analysis reveals that the angular profiles of both the UMR and NPHE are also sensitive to the ratio of the scalar to SOC disorder present in the system, $r \equiv \eta_I / \eta_{\alpha}$ (see Appendix \ref{appendixD} for details). To capture the angular dependencies of the nonlinear responses, combining Eqs.~(\ref{j^2_cubic}) and (\ref{Eq:zeta}), the corresponding UMR coefficients with the magnetization direction being varied in the three orthogonal planes are expressed as
\begin{subequations}
\label{fits_out}
\begin{align}
\left.\zeta_{\parallel}^{(2)}
\right|_{\substack{m_x=0}}
&\simeq
f_{\parallel}(\eta_{ex}, r) \cos \beta_{yz}
+
g_{\parallel}(\eta_{ex}, r) \cos^3 \beta_{yz},
\\
\left. \zeta_{\perp}^{(2)}
\right|_{\substack{m_y=0}}
&\simeq
f_{\perp}(\eta_{ex}, r) \cos \beta_{xz}
+
g_{\perp}(\eta_{ex}, r) \cos^3 \beta_{xz},
\end{align}
\end{subequations}
for the out-of-plane sweeps of the magnetization and
\begin{subequations}
\label{fits_out}
\begin{align}
\left.\zeta_{\parallel}^{(2)}
\right|_{\substack{m_z=0}}
&\simeq
h_{\parallel}(\eta_{ex}, r) \sin \beta_{xy}
+
k_{\parallel}(\eta_{ex}, r) \sin^3 \beta_{xy},
\\
\left. \zeta_{\perp}^{(2)}
\right|_{\substack{m_z=0}}
&\simeq
h_{\perp}(\eta_{ex}, r) \cos \beta_{xy}
+
k_{\perp}(\eta_{ex}, r) \cos^3 \beta_{xy},
\end{align}
\end{subequations}
for the in-plane sweep, where the functions on the right-hand side--which, in general, depend on $\eta_{ex}$ and $r$--are given by
\begin{subequations}
\begin{align}
f_{\parallel}(\eta_{ex}, r)
&=
-\frac{2 \sigma_{\parallel,1}^{(2)}}{\sigma_D}
(1 + \lambda_{\parallel}),
\\
g_{\parallel}(\eta_{ex}, r)
&=
-\frac{2 \sigma_{\parallel,1}^{(2)}}{\sigma_D}
(\kappa_{\parallel} - \lambda_{\parallel}),
\\
h_{\parallel}(\eta_{ex}, r)
&=
-\frac{2 \sigma_{\parallel,1}^{(2)}}{\sigma_D}
(1 + \iota_{\parallel}),
\\
k_{\parallel}(\eta_{ex}, r)
&=
-\frac{2 \sigma_{\parallel,1}^{(2)}}{\sigma_D}
(\kappa_{\parallel} - \iota_{\parallel}),
\end{align}
\end{subequations}
and
\begin{subequations}
\begin{align}
f_{\perp}(\eta_{ex}, r)
&=
-\frac{2 \sigma_{\perp,1}^{(2)}}{\sigma_D}
(1 + \lambda_{\perp}),
\\
g_{\perp}(\eta_{ex}, r)
&=
-\frac{2 \sigma_{\perp,1}^{(2)}}{\sigma_D}
(\iota_{\perp} - \lambda_{\perp}),
\\
h_{\perp}(\eta_{ex}, r)
&=
-\frac{2 \sigma_{\perp,1}^{(2)}}{\sigma_D}
(1 + \kappa_{\perp}),
\\
k_{\perp}(\eta_{ex}, r)
&=
-\frac{2 \sigma_{\perp,1}^{(2)}}{\sigma_D}
(\iota_{\perp} - \kappa_{\perp}).
\end{align}
\end{subequations}
To quantify the deviations of the UMR coefficients from sinusoidal behavior for various sweeps of the magnetization, we introduce the dimensionless ratios $a_{\parallel} \equiv g_{\parallel}/f_{\parallel}$, $b_{\parallel} \equiv k_{\parallel}/h_{\parallel}$, $a_{\perp} \equiv g_{\perp}/f_{\perp}$ and $b_{\perp} \equiv k_{\perp}/h_{\perp}$. The values of these ratios for $\eta_{ex}=0.3$ and $\eta_{ex}=0.6$ are presented in Table~\ref{tab1}, from which it is evident that the cubic terms play an increasingly important--and even dominant--role as the exchange and Fermi energies become comparable. It is worth mentioning that for even higher values of $\eta_{ex}$, as indicated by the blue curves in Fig.~\ref{fig3}, the UMR coefficients become increasingly nonlinear in the magnetization, such that higher-order terms in the magnetization must also be taken into account.

\begin{table}
\caption{Transport coefficients for the cubic contributions to the UMR and NPHE at $r=0.2$ and for different values of $\eta_{ex}$. For the rest of the parameters used, see the caption of Fig.~\ref{fig3}.}
\begin{ruledtabular}
\begin{tabular}{l c c c d}
$\eta_{\text{ex}}$ & $a_{\parallel}$ & $b_{\parallel}$ & $a_{\perp}$ & \multicolumn{1}{c}{$b_{\perp}$}
\\
\hline
0.3 & 0.03 & 0.30 & -0.20 & -0.47
\\
0.6 & 0.09 & 1.92 & -0.59 & -0.85
\\
\end{tabular}
\end{ruledtabular}\label{tab1}
\end{table}

\section{Discussion and Conclusion}

In order to understand the physical origin of the unconventional angular dependencies of the UMR and NPHE, we note that this occurs for larger values of $\eta_{ex}$. In systems where the Fermi energy is fixed, it often suffices to consider terms only to first order in the magnetization, which typically leads to sinusoidal angular dependencies. In TIs, however, the tunability of the Fermi level implies that higher-order terms in the magnetization are expected to play an important role when $\eta_{ex} \sim 1$, thereby necessitating the cubic magnetization terms in Eqs.~(\ref{j^2_cubic}).

This is in contrast to nonlinear magnetotransport phenomena driven by hexagonal warping or particle-hole asymmetry \cite{he2018bilinear, he2019nonlinear}. In the case of warping, the six-fold symmetric deformation of the Fermi contour arises from the addition of a cubic-in-momentum term to the Dirac Hamiltonian, while particle-hole asymmetry adds a $k^2$ term to the Hamiltonian. Thus, both effects are dominant in the limit of high Fermi energy, i.e., when $\eta_{ex} \ll 1$, thereby enabling a simple way to distinguish their contributions from the ones presented in this work.

Another nonlinear transport effect which is stronger in the low Fermi energy limit is that generated by current-induced spin polarization. However, the contribution of this effect to the nonlinear transport can be distinguished from the mechanism predicted here by noting that current-induced spin polarization can only produce a longitudinal quadratic response with no quadratic planar Hall counterpart. This may readily be understood by noting that the emergence of a NPHE requires the magnetization to be parallel to the applied current, which, in turn, is perpendicular to the spin polarization. As a result, the magnetization cannot influence the spin polarization and no quadratic Hall response is generated~\cite{yasuda2016large, langenfeld2016exchange, sterk2019magnon, dyrdal2020spin}.

In addition, in terms of magnitude, the UMR coefficient strengths that we predict here are on the order of $10-100$ nm/V and are thus 1-2 orders of magnitude larger than the bilinear magnetoresistance effect predicted in nonmagnetic TIs \cite{dyrdal2020spin}. This can be attributed to the fact the the proximity-induced exchange interaction is much stronger than the Zeeman interaction for typical experimental magnetic field values, and suggests that TI/FI bilayers are generally a better platform for obtaining sizable nonlinear magnetotransport effects.

Yet another quadratic transport effect that can arise in TI/FI bilayers is the intrinsic nonlinear Hall effect generated by the Berry curvature dipole~\cite{sodemann2015quantum}. However, this relies on an out-of-plane magnetization, in contrast to the NPHE predicted in this work, which only arises when then magnetization has an in-plane component. Furthermore, the absence of a longitudinal UMR generated by the Berry curvature dipole provides an additional means to distinguish it from the effects predicted here.

A final point worth mentioning is that the minimal model we consider here for the Dirac surface states is quite general and is not limited to TI/FI bilayer systems. Thus, it is natural to expect that the nonlinear transport effects predicted in this work should, in principle, also arise in generally magnetized TI sysytems, including intrinsic magnetic TIs \cite{zhang2019topological, wang2020dynamical}, magnetically-doped TIs \cite{liu2009magnetic, abanin2011ordering, nomura2011surface} as well as in single TI layers in the presence of an applied magnetic field.

In conclusion, based on a formal evaluation of nonlinear Kubo formulas in the low-termperature limit, we have predicted a UMR and NPHE in a bilayer comprised of a TI and a FI which require no modification of the Dirac Hamiltonian--but instead--arise solely from extrinsic disorder scattering at the interface. Several key features and unique transport signatures have been identified, which enable the electric and magnetic tuning of the nonlinear magnetoresistance effects. We expect that this work will stimulate further theoretical and experimental studies of quantum transport in the nonlinear response regime, paving the way for potential future quantum spintronic applications.

\section*{Acknowledgments}

This work made use of the High Performance Computing Resource in the Core Facility for Advanced Research Computing at Case Western Reserve University. This work was supported by the College of Arts and Sciences, Case Western Reserve University.

\appendix

\section{Symmetry analysis of nonlinear responses}
\label{appendixA}

In this section, we derive the general form of the nonlinear current, $j^{(2)}_i = \sigma_{ijk} E^j E^k$, to third order in the magnetization for the TI/FI bilayer system described by Eqs.~(\ref{Ham_main}). To this end, consider the magnetization expansion of the quadratic conductivity tensor
\begin{equation}
\sigma^{ijk}
=
\sigma_1^{ijkl}m_l
+
\sigma_2^{ijklm}m_l m_m
+
\sigma_3^{ijklmn}m_l m_m m_n.
\end{equation}
Without loss of generality, let us set the electric field in the $\mathbf{x}$ direction so that we need only evaluate $\sigma^{xxx}$ for the longitudinal current. Note that in the absence of the exchange interaction, the system is even under the mirror reflection transformation in the $xz$ ($yz$) plane, $\mathcal{M}_{xz}$ ($\mathcal{M}_{yz}$), and thus so are $\sigma_1$, $\sigma_2$ and $\sigma_3$. Using the fact that the current and magnetization are polar and axial vectors, respectively, it is then straightforward to verify that, upon imposing $\mathcal{M}_{xz}$ and $\mathcal{M}_{yz}$, $\sigma_1^{xxxx}=\sigma_1^{xxxz}=0$, while, in general, $\sigma_1^{xxxy} \neq 0$. Similarly, for the quadratic contribution, one can verify that only terms $\propto m_x m_z$ are allowed by reflection symmetry. However, given the particle-hole symmetry and the quadratic dependence of the disorder averages on $m_z$ [see Eqs.~(\ref{vvavg})], the terms linear in $m_z$ do not contribute to the nonlinear response and $\sigma_2$, therefore, plays no role in the nonlinear transport.

Moving on to the cubic contribution, one can verify that, in addition to $\sigma_3^{xxxxyz}$, terms containing even powers of $m_y$ do not contribute to $j_x^{(2)}$. Thus, of the 10 independent components of $\sigma_3^{xxxijk}$, only 3 survive the symmetry requirements, namely $\sigma_3^{xxxyxx}$, $\sigma_3^{xxxyyy}$ and $\sigma_3^{xxxyzz}$. By an analogous argument for the transverse current, one can show that only the  independent components $\sigma_1^{yxxx}$, $\sigma_3^{yxxxxx}$, $\sigma_3^{yxxxyy}$ and $\sigma_3^{yxxxzz}$ may be nonvanishing. Therefore, the nonlinear current up to third order in the magnetization may be expressed as
\begin{subequations}
\begin{align}
j_x^{(2)}
&=
j_{x,1}^{(2)} 
\left(
1 + \iota_{\parallel} m_x^2 + \kappa_{\parallel} m_y^2 + \lambda_{\parallel} m_z^2
\right),
\\
j_y^{(2)}
&=
j_{y,1}^{(2)} 
\left(
1 + \iota_{\perp} m_x^2 + \kappa_{\perp} m_y^2 + \lambda_{\perp} m_z^2
\right),
\end{align}
\end{subequations}
where
$j_{x,1}^{(2)}
=
\sigma_1^{xxxy} m_y E_x^2$, 
$j_{y,1}^{(2)}
=
\sigma_1^{yxxx} m_x E_x^2$ and the dimensionless functions measuring the cubic contributions read
\begin{subequations}
\begin{align}
\iota_{\parallel}
&=
\frac{3 \sigma_3^{xxxyxx}}{\sigma_1^{xxxy}},
\\
\kappa_{\parallel}
&=
\frac{\sigma_3^{xxxyyy}}{\sigma_1^{xxxy}},
\\
\lambda_{\parallel}
&=
\frac{3 \sigma_3^{xxxyzz}}{\sigma_1^{xxxy}},
\end{align}
\end{subequations}
and
\begin{subequations}
\begin{align}
\iota_{\perp}
&=
\frac{\sigma_3^{yxxxxx}}{\sigma_1^{yxxx}},
\\
\kappa_{\perp}
&=
\frac{3 \sigma_3^{yxxxyy}}{\sigma_1^{yxxx}},
\\
\lambda_{\perp}
&=
\frac{3 \sigma_3^{yxxxzz}}{\sigma_1^{yxxx}}.
\end{align}
\end{subequations}
Relaxing the orientation of the electric field to point in an arbitrary direction in the $xy$ plane, we arrive at Eqs.~(\ref{j^2_cubic}), where  the functions $\iota$, $\kappa$ and $\lambda$ are now understood to be evaluated with the relevant tensor components in the basis spanned by $\mathbf{e}$, $\mathbf{z} \times \mathbf{e}$ and $\mathbf{z}$. 

\section{Self-energy and scattering time}
\label{appendixB}

The unperturbed Hamiltonian, Eq.~(\ref{H0_main}), has eigenstates given by
\begin{equation}
\ket{u_{\mathbf{q}\sigma}}
=
\begin{pmatrix}
\left( \frac{1+\sigma}{2}\right) 
\cos \frac{\theta_{\mathbf{h}}}{2}
+
\left( \frac{1-\sigma}{2}\right) 
\sin \frac{\theta_{\mathbf{h}}}{2}
\\
e^{i\phi_\mathbf{h}}
\left[
\left( \frac{1+\sigma}{2}\right) 
\sin \frac{\theta_{\mathbf{h}}}{2}
-
\left( \frac{1-\sigma}{2}\right) 
\cos \frac{\theta_{\mathbf{h}}}{2}
\right]
\end{pmatrix},
\end{equation}
where $\phi_\mathbf{h}$ is the azimuthal angle in the plane spanned by $\mathbf{h}_{\mathbf{q}}$, with $\cos \phi_\mathbf{h} = h_{x\mathbf{q}}/h_{\mathbf{q}}$ and $\sin \phi_\mathbf{h} = h_{y\mathbf{q}}/h_{\mathbf{q}}$, while $\theta_{\mathbf{h}}$ is the polar angle, with $\sin \theta_\mathbf{h} = h_{\mathbf{q}}/\sqrt{h_{\mathbf{q}}^2 + \Delta_{ex}^2 m_z^2}$. The velocity operator is defined as $\hat{\mathbf{v}}_{\mathbf{q}}=\bs{\pd}_{\mathbf{q}}\hat{H}_{\mathbf{q}}^0/\hbar$, with $\pd_{\mathbf{q}}^i \equiv \pd/\pd q_i$, which in the chiral basis of Bloch eigenstates, leads to the diagonal terms $\mathbf{v}_{\mathbf{q}\sigma}=\pd_{\mathbf{q}}\epsilon_{\mathbf{q}\sigma}/\hbar$, or
\begin{subequations}
\label{S_v_qs}
\begin{align}
v_{\mathbf{q}\sigma}^x
&=
- \sigma v_F \sin \theta_\mathbf{h} \sin \phi_{\mathbf{h}},
\\
v_{\mathbf{q}\sigma}^y
&=
\sigma v_F \sin \theta_\mathbf{h} \cos \phi_{\mathbf{h}}.
\end{align}
\end{subequations}
In the chiral Bloch basis the disorder averages read
\begin{widetext}
\begin{subequations}
\label{vvavg}
\begin{align}
\Braket{V_{\mathbf{q}\mathbf{q}^{\prime}}^{\sigma \sigma^{\prime}} V_{\mathbf{q}^{\prime}\mathbf{q}}^{{\sigma^{\prime} \sigma}}}
&=
\Braket{U_{\mathbf{q}\mathbf{q}^{\prime}}^{\sigma \sigma^{\prime}} U_{\mathbf{q}^{\prime}\mathbf{q}}^{{\sigma^{\prime} \sigma}}}
+
\Braket{W_{\mathbf{q}\mathbf{q}^{\prime}}^{\sigma \sigma^{\prime}} W_{\mathbf{q}^{\prime}\mathbf{q}}^{{\sigma^{\prime} \sigma}}},
\\
\Braket{U_{\mathbf{q}\mathbf{q}^{\prime}}^{\sigma \sigma^{\prime}} U_{\mathbf{q}^{\prime}\mathbf{q}}^{{\sigma^{\prime} \sigma}}}
&=
\frac{1}{2} n_I U_0^2
\left[ 1 
+
\sigma \sigma^{\prime} 
\cos \theta_{\mathbf{h}} \cos \theta_{\mathbf{h}^{\prime}}
+
\sigma \sigma^{\prime} 
\cos \left(\phi_{\mathbf{h}} - \phi_{\mathbf{h}^{\prime}}\right)
\sin \theta_{\mathbf{h}} \sin \theta_{\mathbf{h}^{\prime}}\right],
\\
\begin{split}
\Braket{W_{\mathbf{q}\mathbf{q}^{\prime}}^{\sigma \sigma^{\prime}} W_{\mathbf{q}^{\prime}\mathbf{q}}^{{\sigma^{\prime} \sigma}}}
&=
\frac{1}{8} n_{\alpha} W_0^2
\left\{
\left[ 1 
-
\sigma \sigma^{\prime} 
\cos \theta_{\mathbf{h}} \cos \theta_{\mathbf{h}^{\prime}}\right]
\left[ \left(q_x + q_x^{\prime}\right)^2
+ \left(q_y + q_y^{\prime}\right)^2 \right]
\right.
\\
&\left. 
-
\sigma \sigma^{\prime} 
\cos \left(\phi_{\mathbf{h}} + \phi_{\mathbf{h}^{\prime}}\right)
\sin \theta_{\mathbf{h}} \sin \theta_{\mathbf{h}^{\prime}}
\left[ \left(q_x + q_x^{\prime}\right)^2
- \left(q_y + q_y^{\prime}\right)^2 \right]
\right.
\\
&\left. 
-
2 \sigma \sigma^{\prime} 
\sin \left(\phi_{\mathbf{h}} + \phi_{\mathbf{h}^{\prime}}\right)
\sin \theta_{\mathbf{h}} \sin \theta_{\mathbf{h}^{\prime}}
\left(q_x + q_x^{\prime}\right)
\left(q_y + q_y^{\prime}\right)
\right\},
\end{split}
\end{align}
\end{subequations}
where $V_{\mathbf{q}\mathbf{q}^{\prime}}^{\sigma \sigma^{\prime}} 
\equiv
\braket{u_{\mathbf{q}\sigma}| \hat{V}_{\mathbf{q}\mathbf{q}^{\prime}} |u_{\mathbf{q}^{\prime}\sigma^{\prime}}}$.
Inserting Eqs. (\ref{vvavg}) into Eq.~(\ref{selfenergy}), we obtain
\begin{equation}
\Gamma_{\mathbf{q}\sigma}
=
\Gamma_I
\left[1+ \sigma \eta_{\text{ex}} m_z \cos \theta_{\mathbf{h}}
+
\frac{\Gamma_{\alpha}}{\Gamma_I} 
\mathcal{F}_{\sigma}
\left( \bs{\eta}_{\mathbf{h}}, \eta_{ex} ; \mathbf{m}\right)\right],
\end{equation}
where the dimensionless function $\mathcal{F}_{\sigma}$ reads
\begin{equation}
\begin{split}
\mathcal{F}_{\sigma}
\left( \bs{\eta}_{\mathbf{h}}, \eta_{ex} ; \mathbf{m} \right)
&=
\left(1-\sigma \eta _{ex} m_z \cos \theta_{\mathbf{h}} \right)
\left[1 + \eta_{\mathbf{h}}^2 + \eta _{ex}^2 \left(4 - 5 m_z^2\right)
-
4 \eta_{ex} \bs{\eta}_{\mathbf{h}} \cdot \mathbf{m} \right]
\\
&+
2 \sigma \sin \theta_{\mathbf{h}}
\left(1- \eta_{ex}^2 m_z^2 \right)
\left( \eta _{\mathbf{h}} - 2 \eta_{ex} \frac{\bs{\eta}_{\mathbf{h}} \cdot \mathbf{m} }{\eta_{\mathbf{h}}}\right),
\end{split}
\end{equation}
in which we assume the Fermi level lies in the upper band ($\epsilon>0$). From this, we arrive at the momentum scattering time, Eq.~(\ref{tau_qsigma})

\section{Vertex correction and nonlinear conductivity}
\label{appendixC}

The disorder-averaged velocity vertex function may be found self-consistently from the general vertex equation
\begin{equation}
\bs{\mathcal{V}}_{\mathbf{q}\sigma} \left(\epsilon, \epsilon^{\prime}\right)
=
\mathbf{v}_{\mathbf{q}\sigma}
+
\sum_{\mathbf{q}^{\prime}\sigma^{\prime}}
\Braket{V_{\mathbf{q}\mathbf{q}^{\prime}}^{\sigma \sigma^{\prime}} V_{\mathbf{q}^{\prime}\mathbf{q}}^{{\sigma^{\prime} \sigma}}}
G^{\text{A/R}}_{\mathbf{q}^{\prime}\sigma^{\prime}}\left(\epsilon\right)
\bs{\mathcal{V}}_{\mathbf{q}^{\prime}\sigma^{\prime}} \left(\epsilon, \epsilon^{\prime}\right)
G^{\text{R/A}}_{\mathbf{q}^{\prime}\sigma^{\prime}}\left(\epsilon^{\prime}\right),
\end{equation}
where $\epsilon$ and $\epsilon^{\prime}$ are, respectively, the energies of the incoming and outgoing Green's functions to the vertex in question. The near-dc behavior may be captured by setting $\epsilon^{\prime}=\epsilon+ \hbar \omega$. Then, using the identity
\begin{equation}
G^{\text{A/R}}_{\mathbf{q}\sigma}
\left( \epsilon \right)
G^{\text{R/A}}_{\mathbf{q}\sigma}
\left( \epsilon + \hbar \omega \right)
\simeq
\frac{2\pi}{\hbar} \frac{\tau_{\mathbf{q}\sigma}}{1 \mp i \omega \tau_{\mathbf{q}\sigma}}
\delta \left( \epsilon_{\mathbf{q}\sigma} - \epsilon \right),
\end{equation}
the vertex function may be approximated as
\begin{equation}
\label{V_approx}
\bs{\mathcal{V}}_{\mathbf{q}\sigma} \left(\epsilon, \epsilon + \hbar \omega\right)
\simeq
\mathbf{v}_{\mathbf{q}\sigma}
+
\frac{2\pi}{\hbar}
\sum_{\mathbf{q}^{\prime}\sigma^{\prime}}
\Braket{V_{\mathbf{q}\mathbf{q}^{\prime}}^{\sigma \sigma^{\prime}} V_{\mathbf{q}^{\prime}\mathbf{q}}^{{\sigma^{\prime} \sigma}}}
\frac{\tau_{\mathbf{q}^{\prime}\sigma^{\prime}}}{1 \mp i \omega \tau_{\mathbf{q}^{\prime}\sigma^{\prime}}}
\mathbf{v}_{\mathbf{q}^{\prime}\sigma^{\prime}} 
\delta \left( \epsilon_{\mathbf{q}^{\prime}\sigma^{\prime}} - \epsilon \right).
\end{equation}
To obtain an approximate form for the conductivity tensor, we insert Eq.~(\ref{V_approx}) into Eq.~(\ref{sigma_ijk}) and use the identity
\begin{equation}
\label{iden_cond}
\left[G^{\text{R}}_{\mathbf{q}\sigma}
\left( \epsilon \right)\right]^{m+1}
\left[G^{\text{A}}_{\mathbf{q}\sigma}
\left( \epsilon \right)\right]^{n+1}
\simeq
2\pi i^{n-m}\frac{\left(m+n\right)!}{m!n!} \left(\frac{\tau_{\mathbf{q}\sigma}}{\hbar}\right)^{m+n+1}
\delta \left( \epsilon_{\mathbf{q}\sigma} - \epsilon \right),
\end{equation}
to arrive at the following form for the in-plane conductivity components
\begin{equation}
\label{S_sigma_ijj}
\sigma^{ijj}
=
- \frac{4 e^3}{\hbar}
\text{Re} \sum_{\mathbf{q}}
\left[
\pd_{\omega}
\mathcal{V}^i_{\mathbf{q} +} \left(\epsilon_F, \epsilon_F + \hbar \omega \right) \Bigr|_{\substack{\omega=0}}
+
i \mathcal{V}_{\mathbf{q} +}^{i F} \tau_{\mathbf{q} +}^F
\left(
1 + i \frac{\pd \Gamma_{\mathbf{q} +}^F}{\pd \epsilon_F}
\right)
\right]
v_{\mathbf{q} +}^j \mathcal{V}_{\mathbf{q} +}^{j F} \left( \tau_{\mathbf{q} +}^F \right)^2 \delta \left( \epsilon_{\mathbf{q} +} - \epsilon_F \right),
\end{equation}
where $\Gamma_{\mathbf{q} \sigma}^F$ and $\tau_{\mathbf{q} \sigma}^F$ are the self-energy and scattering time at the Fermi level. In this form, the quadratic conductivity tensor is expressed entirely in terms of the velocity vertex function and scattering time at the Fermi level. Plots of the angular dependencies of these two quantities for various sweeps of the magnetization are presented in Fig.~\ref{figS1}.

\begin{figure}[tph]
    \sidesubfloat[]{\includegraphics[width=0.2\linewidth,trim={1.5cm .3cm .5cm 1cm}]{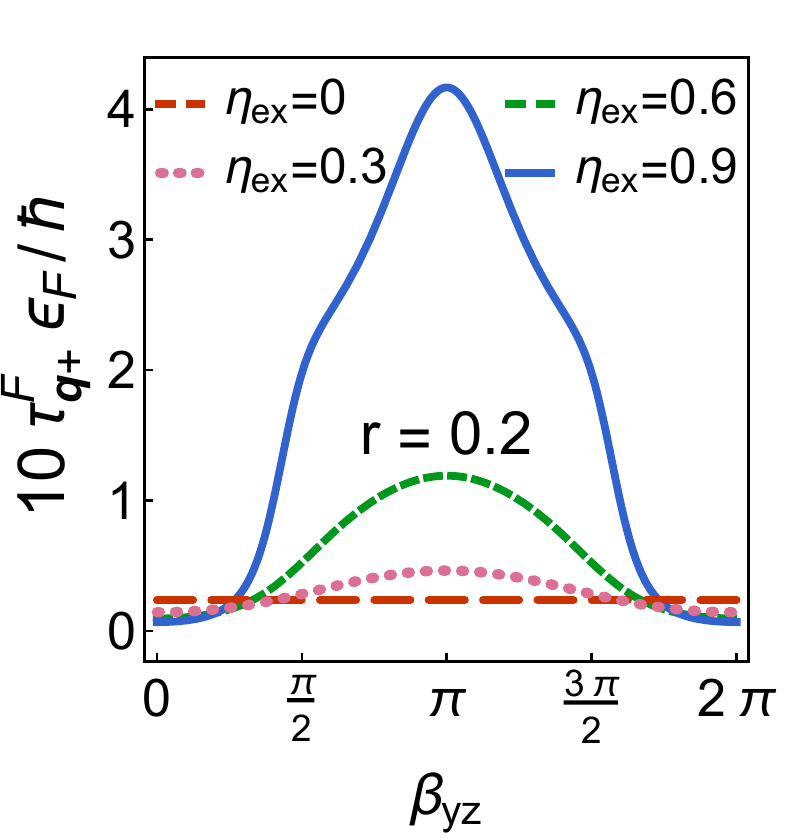}\label{fig4a}}
    \sidesubfloat[]{\includegraphics[width=0.2\linewidth,trim={1.1cm .2cm .5cm .8cm}]{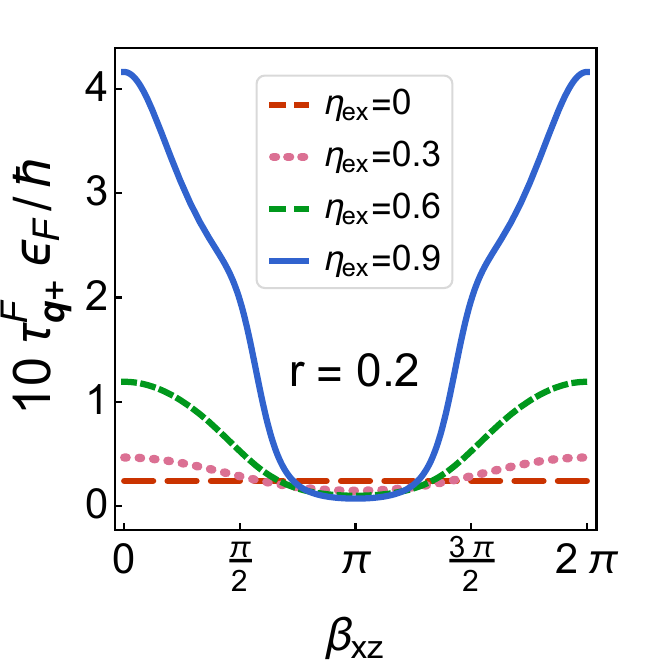}\label{fig4b}}
    \\
    \sidesubfloat[]{\includegraphics[width=0.2\linewidth,trim={1.9cm .9cm .8cm .8cm}]{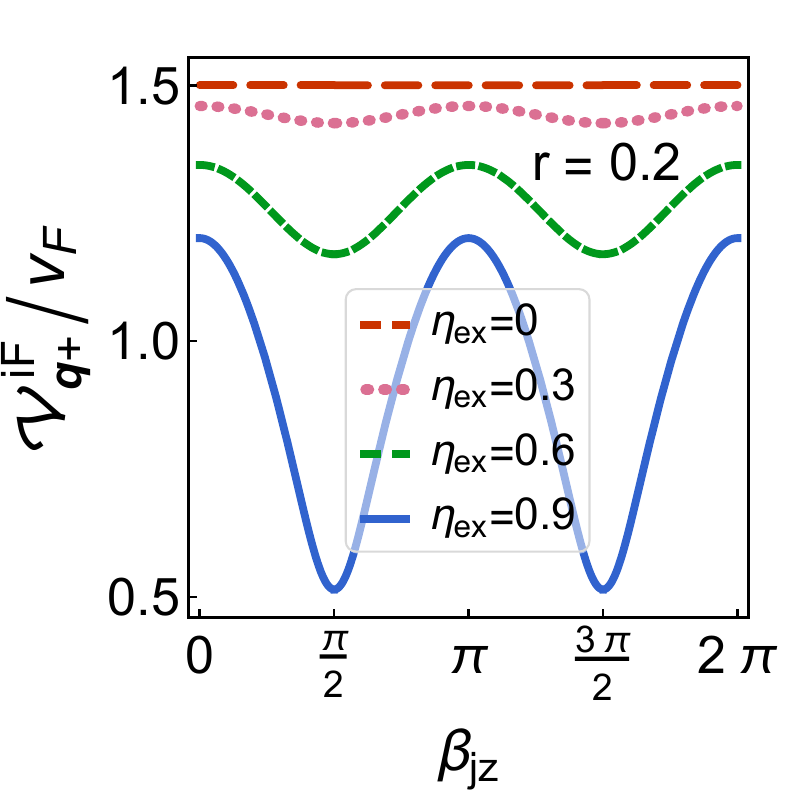}\label{fig4c}}
    \sidesubfloat[]{\includegraphics[width=0.2\linewidth,trim={1.3cm .9cm .7cm .7cm}]{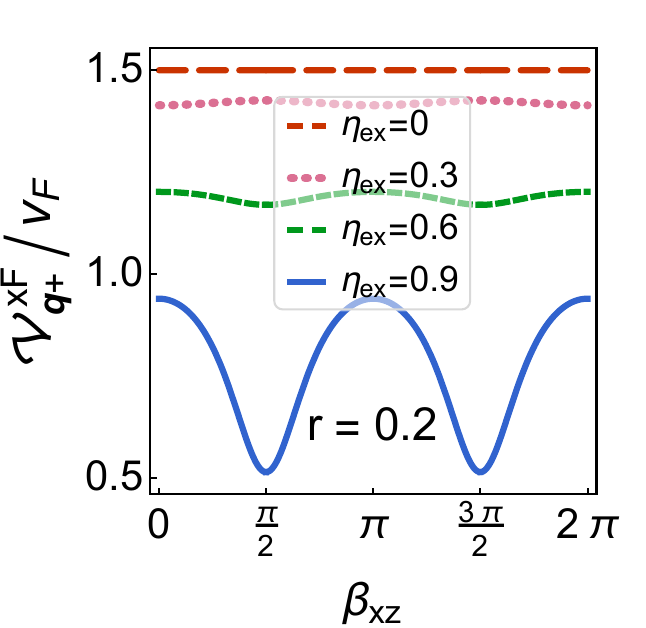}\label{fig4d}}    
    \caption{Angular dependencies at the Fermi level.  (a) scattering time (with $q_y=0$) in the $yz$ plane, (b) $i$-th component of the velocity vertex function (with $q_j=0$) in the $jz$ plane, where $i/j = x/y$ or $y/x$. (c) scattering time (with $q_x=0$) in the $xz$ plane and (d) longitudinal velocity vertex function (with $q_y=0$) in the $xz$ plane. Parameters used: $\epsilon_F=0.5$ eV, $v_F= 5\times10^{14}$ nm/s \cite{qi2011topological} and $\eta_I = 0.01$.}
    \label{figS1}
\end{figure}

\section{Angular dependencies with weak SOC disorder}
\label{appendixD}

Here, we analyze the angular dependencies of the UMR coefficients in the limit where the SOC disorder is weaker than the scalar disorder. We thus set $r=2$ and plot the UMR coefficients, which are presented in Fig.~\ref{figS2}. The first notable feature is that compared to the plots in Fig.~\ref{fig3}--wherein $r=0.2$--the amplitudes of the UMR and NPHE are generally weaker here for various in-plane and out-of-plane magnetization sweeps. This is not surprising, as the effects are generated in the presence of SOC disorder, which is also weaker here.

Another interesting feature is the qualitative behavior of the UMR coefficients when the Fermi level approaches the exchange energy. For out-of-plane sweeps of the magnetization, as shown in Figs.~\ref{figS1a} and \ref{figS1b}, for lower values of $\eta_{ex}$ (see the dotted pink and dashed green curves), the UMR (NPHE) displays typical sinusoidal behavior, with 2 symmetry-imposed sign changes at $\beta_{yz}$($\beta_{xz}$) $= \pi/2, 3\pi/2$. However, at $\eta_{ex}=0.9$ (see the solid blue curves), 4 additional sign change points appear, which are connected by quasiplateaus of suppressed conductivity.

In order to gain further insight into the emergence of quasiplateaus and additional sign change points in the nonlinear magnetoresistances, it is convenient to obtain an approximate analytical expression for the conductivity tensor. As the irregularities in the angular dependencies emerge when the magnetization is almost out of plane, we may treat the in-plane components of the unit magnetization vector as perturbative parameters in the quasiplateau-forming angular region. Inserting Eqs.~(\ref{S_v_qs}) and (\ref{vvavg}) into Eq.~(\ref{V_approx}), to first order in $m_x$ and $m_y$, the dressed velocity vertex function may be expressed as
\begin{equation}
\label{vertex_soln}
\begin{split}
\bs{\mathcal{V}}_{\mathbf{q}\sigma} \left(\epsilon, \epsilon + \hbar \omega\right)
&=
\mathcal{I}_{\sigma}\left( \bs{\eta}_{\mathbf{h}}, \eta_{ex}, m_z , \Gamma_{\alpha}, \Gamma_{I}; \omega \right)
\mathbf{v}_{\mathbf{q}\sigma}
\\
&+
v_F \mathbf{z} \times
\left[ \mathcal{J}_{\sigma}\left( \bs{\eta}_{\mathbf{h}}, \eta_{ex}, m_z , \Gamma_{\alpha}, \Gamma_{I}; \omega \right)
\bs{\eta}_{\mathbf{h}}
+
\eta_{ex} \mathcal{K}_{\sigma}\left( \bs{\eta}_{\mathbf{h}}, \eta_{ex}, m_z , \Gamma_{\alpha}, \Gamma_{I}; \omega \right)
\mathbf{m}
\right],
\end{split}
\end{equation}
where the dimensionless functions $\mathcal{I}_{\sigma}$, $\mathcal{J}_{\sigma}$ and $\mathcal{K}_{\sigma}$ are given by

\begin{subequations}
\begin{align}
\begin{split}
\mathcal{I}_{\sigma}\left( \bs{\eta}_{\mathbf{h}}, \eta_{ex}, m_z , \Gamma_{\alpha}, \Gamma_{I}; \omega \right)
&=
\frac{\Gamma _{\alpha} \left(1-\eta _{\text{ex}}^2 m_z^2 \right) \left[ 9 \left( 1 - \eta _{\text{ex}}^2 m_z^2 \right) + \eta _{\mathbf{h}}^2 \right]+\Gamma _I \left( 3 + \eta _{\text{ex}}^2 m_z^2 \right) \mp i \hbar \omega}
{8 \Gamma _{\alpha} \left(1 - \eta _{\text{ex}}^2 m_z^2 \right)^2 + 2 \Gamma _I \left( 1+ \eta _{\text{ex}}^2 m_z^2 \right) \mp i \hbar \omega},
\end{split}
\\
\begin{split}
\mathcal{J}_{\sigma}\left( \bs{\eta}_{\mathbf{h}}, \eta_{ex}, m_z , \Gamma_{\alpha}, \Gamma_{I}; \omega \right)
&=
\frac{2 \Gamma _{\alpha} 
\left( 1 - \eta _{\text{ex}}^2 m_z^2 \right) 
\left( 1-\sigma  \eta _{\text{ex}} m_z \cos \theta_{\mathbf{h}} \right)}
{8 \Gamma _{\alpha } \left(1-\eta _{\text{ex}}^2 m_z^2 \right)^2 + 2 \Gamma _I \left( 1 + \eta _{\text{ex}}^2 m_z^2 \right) \mp i \hbar \omega},
\end{split}
\\
\begin{split}
\mathcal{K}_{\sigma}\left( \bs{\eta}_{\mathbf{h}}, \eta_{ex}, m_z , \Gamma_{\alpha}, \Gamma_{I}; \omega \right)
&=
- \frac{4 \Gamma _{\alpha}
\left(1-\eta _{\text{ex}}^2 m_z^2 \right)}
{\left[ 8 \Gamma _{\alpha} 
\left( 1-\eta _{\text{ex}}^2 m_z^2 \right)^2 + 2 \Gamma _I \left(1 + \eta _{\text{ex}}^2 m_z^2 \right) \mp i \hbar \omega \right]^2}
\\
&\hspace{0.025\linewidth}\times
\Big\{
4 \Gamma _{\alpha}
\left( 1-\eta_{\text{ex}}^2 m_z^2 \right) 
\left( 1 - \eta _{\text{ex}}^2 m_z^2 - \eta_{\mathbf{h}}^2 \right)
\left( 1 - \sigma \eta _{\text{ex}} m_z \cos \theta_{\mathbf{h}} \right)
\\
&\hspace{0.025\linewidth}
- 2 \Gamma_I
\left[ 1- 3 \eta _{\text{ex}}^2 m_z^2 + \sigma \eta _{\text{ex}} m_z \left(3-\eta _{\text{ex}}^2 m_z^2 \right) 
\cos \theta_{\mathbf{h}}
- \sigma \eta_{\mathbf{h}} \left( 1 + \eta _{\text{ex}}^2 m_z^2 \right) \sin \theta_{\mathbf{h}}
\right]
\\
&\hspace{0.025\linewidth}
\mp i \hbar \omega \left[
1 - \sigma \eta _{\text{ex}} m_z \cos \theta_{\mathbf{h}} + \sigma  \eta_{\mathbf{h}} \sin \theta_{\mathbf{h}}
\right]
\Big\}.
\end{split}
\end{align}
\end{subequations}

Inserting this solution into Eq.~(\ref{S_sigma_ijj}), the longitudinal and transverse conductivities read
\begin{subequations}
\label{sigma_anal}
\begin{align}
\sigma_{xxx}
&=
-\frac{9 e^3 v_F}{32 \pi \epsilon_F^2} m_y
\,\mathcal{C} \left( \eta_{\alpha}, \eta_{I}, \eta_{ex}, m_z \right),
\\
\sigma_{yxx}
&=
\frac{3 e^3 v_F}{32 \pi \epsilon_F^2} m_x
\,\mathcal{C} \left( \eta_{\alpha}, \eta_{I}, \eta_{ex}, m_z \right),
\end{align}
\end{subequations}
where $\eta_{ex}$ is understood to be calculated at the Fermi level and the dimensionless function $\mathcal{C}$ is given by
\begin{equation}
\label{C_fun}
\begin{split}
\mathcal{C} \left( \eta_{\alpha}, \eta_{I}, \eta_{ex}, m_z \right)
&=
\eta _{\alpha} \eta_{ex}
\left( 1 - \eta _{\text{ex}}^2 m_z^2 \right)^2
\left[12 \eta_{\alpha}
\left( 1 - \eta _{\text{ex}}^2 m_z^2 \right)^2
+ \eta_I
\left( 1 - 3 \eta _{\text{ex}}^2 m_z^2 \right) \right] 
\\
&\times
\left[ 12 \eta_{\alpha}
\left( 1 - \eta _{\text{ex}}^2 m_z^2 \right)^2
+ \eta _I
\left( 3 + \eta _{\text{ex}}^2 m_z^2 \right) \right]^2
\left[ 4 \eta_{\alpha}
\left( 1 - \eta _{\text{ex}}^2 m_z^2 \right)^2
+ \eta_I
\left( 1 + \eta _{\text{ex}}^2 m_z^2 \right) \right]^{-6},
\end{split}
\end{equation}

\begin{figure}[hpt]
    \sidesubfloat[]{\includegraphics[width=0.2\linewidth,trim={.7cm 0cm .7cm 0.5cm}]{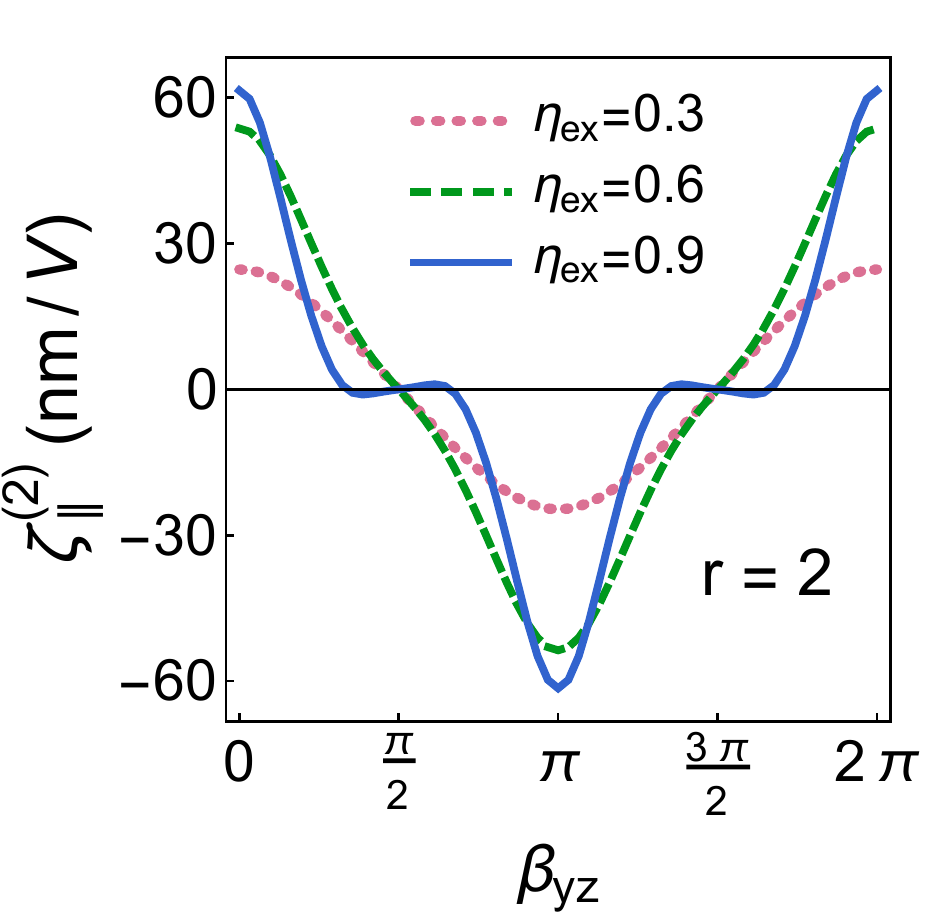}\label{figS1a}}
    \sidesubfloat[]{\includegraphics[width=0.2\linewidth,trim={.7cm 0cm .7cm .5cm}]{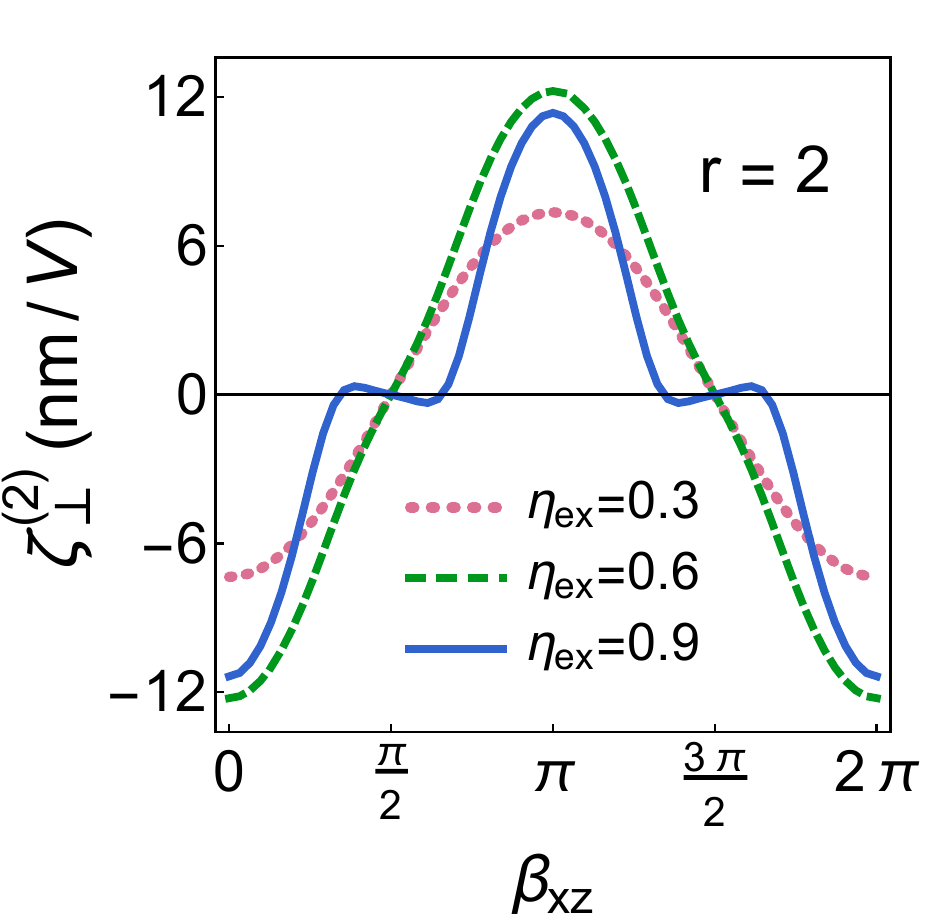}\label{figS1b}}
\\
    \sidesubfloat[]{\includegraphics[width=0.2\linewidth,trim={.7cm .7cm .7cm .5cm}]{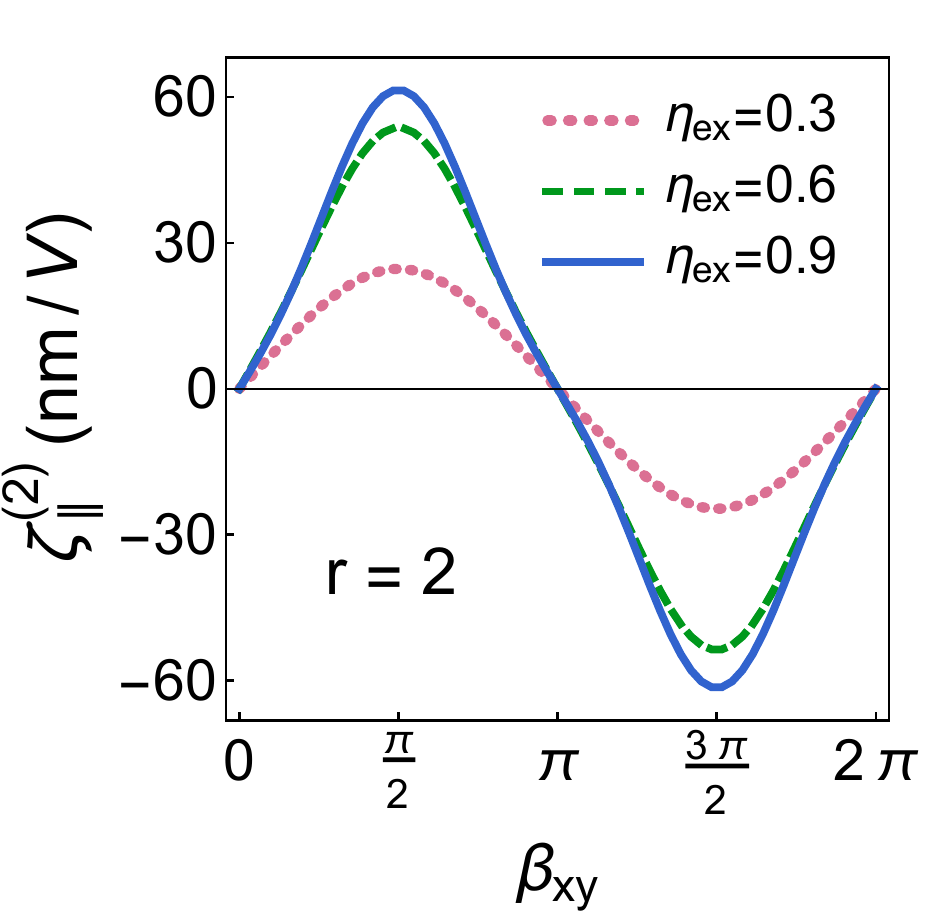}\label{figS1c}}
    \sidesubfloat[]{\includegraphics[width=0.2\linewidth,trim={.7cm .7cm .7cm .5cm}]{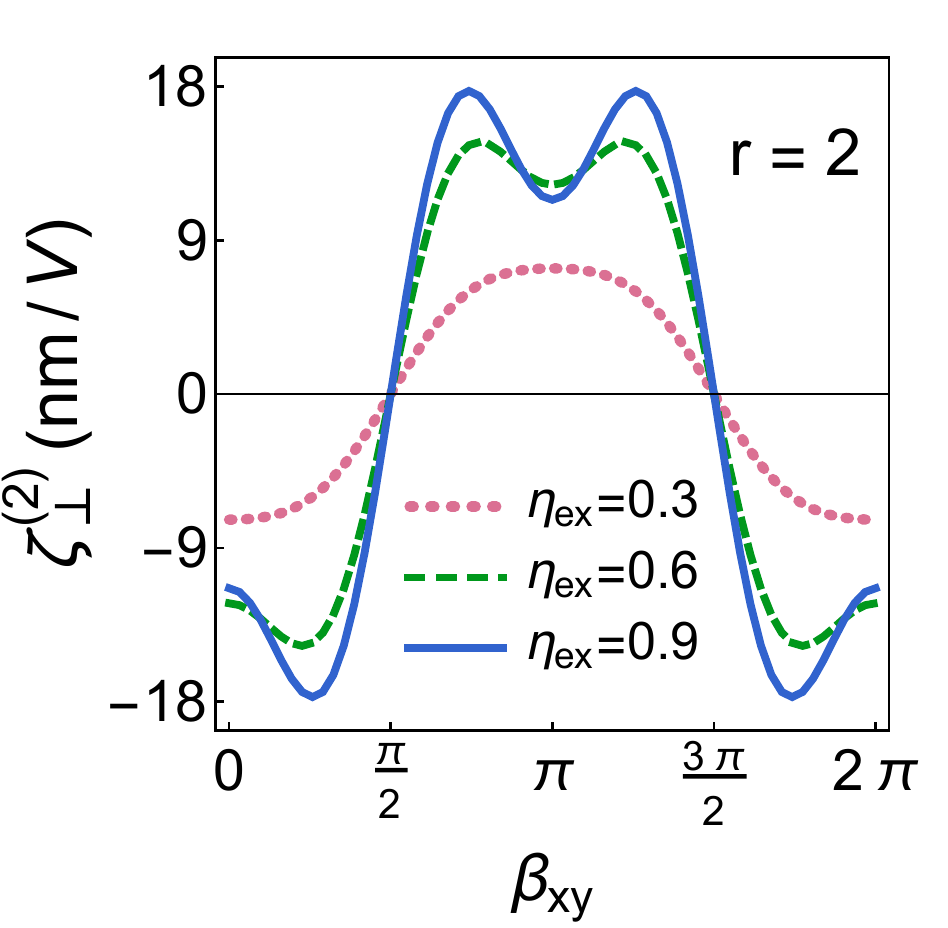}\label{figS1d}}
    \caption{Angular dependencies of the UMR coefficients for various angular sweeps of the magnetization. Here, $\beta_{ij}$ ($i,j=x,y,z$) is the angle between the $i$ axis and the magnetization as it sweeps the $ij$ plane. Parameters used: $\epsilon_F=0.5$ eV, $v_F= 5\times10^{14}$ nm/s \cite{qi2011topological} and $\eta_I = 0.01$.}
        \label{figS2}
\end{figure}
leading to the UMR coefficients
\begin{subequations}
\label{zeta_anal}
\begin{align}
\zeta _{\parallel }^{(2)}
&=
\frac{9 e^3 v_F}{16 \pi \sigma_D \epsilon_F^2} m_y
\,\mathcal{C} \left( \eta_{\alpha}, \eta_{I}, \eta_{ex}, m_z \right),
\\
\zeta _{\perp}^{(2)}
&=
-\frac{3 e^3 v_F}{16 \pi \sigma_D \epsilon_F^2} m_x
\,\mathcal{C} \left( \eta_{\alpha}, \eta_{I}, \eta_{ex}, m_z \right).
\end{align}
\end{subequations}
Note that in this approximation, the ratio between the longitudinal and transverse UMR coefficients takes the simple form
\begin{equation}
\frac{\zeta _{\perp}^{(2)}}{\zeta _{\parallel}^{(2)}}
=
- \frac{1}{3} \cot \phi_{\mathbf{m}},
\end{equation}
where $\phi_{\mathbf{m}}$ is the angle between the electric field and magnetization, further confirming the common physical origin of the two nonlinear magnetoresistances.

Figs~\ref{figS1a} and \ref{figS1b} reveal that when an out-of-plane angular sweep of the magnetization is performed, there will be a total of 6 angles where the longitudinal or transverse UMR coefficients vanish. Two of these are the conventional angles where the in-plane magnetization vanishes ($\beta_{yz}= \pi/2, 3\pi/2$ for the UMR and $\beta_{xz}= \pi/2, 3\pi/2$ for the NPHE). The other four must than correspond to solutions of $\mathcal{C}=0$, which turn out to be $\beta_{yz}= \pi/2 \pm \theta_p/2, 3\pi/2 \pm \theta_p/2$, and similarly for $\beta_{xz}$, where $\theta_p$ is the angular arc over which the conductivities display a plateau-like profile and is given by $\theta_p = 2 \cos^{-1}[f(r)/\eta_{ex}]$, where $r=\eta_I/ \eta_{\alpha}$ and the dimensionless function $f(r)$ is
\begin{equation}
\label{theta_p}
f(r)
=
\sqrt{1 + \frac{r}{8}
\left(1- \sqrt{1+ \frac{32}{3r}}\right)}.
\end{equation}
\end{widetext}

\begin{figure}[tph]
{\includegraphics[width=0.8\linewidth,trim={0cm 0 0cm -1cm}]{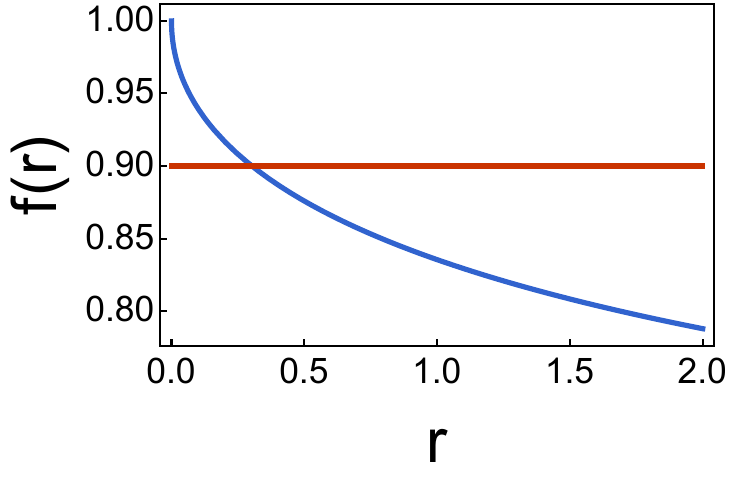}}%
    \caption{Plot of the monotonically decreasing function $f(r)$ (blue), with the red horizontal line indicating the value of $\eta_{ex}=0.9$ for comparison. The unconventional angular dependencies of the conductivities emerge when $\eta_{ex}>f(r)$.}
    \label{figS3}
\end{figure}

For the parameters used in obtaining Figs~\ref{figS1a} and \ref{figS1b}, $\eta_{ex}=0.9$ and $r=2$, we find that $\theta_p \approx \pi/3$, which is quite sizable. From Eq. (\ref{theta_p}), we may deduce the requirement the system imposes for the additional sign changes of the nonlinear conductivity--and therefore the emergence of the quasiplateaus--to occur. From the condition that $\theta_p>0$, we conclude that the system must satisfy $\eta_{ex}>f(r)$. This requirement--or its lack thereof--is met by all the curves in Figs.~\ref{fig3a} and \ref{fig3c} as well as the ones presented in Fig.~\ref{figS2}, further confirming the validity of the analytical approximation. In addition, as displayed in Fig.~\ref{figS3}, $f(r)$ is a monotonically decreasing function. From this fact, and the requirement that $\eta_{ex}>f(r)$, we conclude that in order for the system to allow for additional sign changes of the nonlinear magnetoresistances and the subsequent emergence of plateau-like profiles, the Fermi level must be sufficiently close to the exchange energy and the density of the scalar scatterers must be sufficiently higher than that of the SOC impurities.

\bibliographystyle{my-aps-style}
\bibliography{proximity}

\begin{thebibliography}{68}%
\makeatletter
\providecommand \@ifxundefined [1]{%
 \@ifx{#1\undefined}
}%
\providecommand \@ifnum [1]{%
 \ifnum #1\expandafter \@firstoftwo
 \else \expandafter \@secondoftwo
 \fi
}%
\providecommand \@ifx [1]{%
 \ifx #1\expandafter \@firstoftwo
 \else \expandafter \@secondoftwo
 \fi
}%
\providecommand \natexlab [1]{#1}%
\providecommand \enquote  [1]{``#1''}%
\providecommand \bibnamefont  [1]{#1}%
\providecommand \bibfnamefont [1]{#1}%
\providecommand \citenamefont [1]{#1}%
\providecommand \href@noop [0]{\@secondoftwo}%
\providecommand \href [0]{\begingroup \@sanitize@url \@href}%
\providecommand \@href[1]{\@@startlink{#1}\@@href}%
\providecommand \@@href[1]{\endgroup#1\@@endlink}%
\providecommand \@sanitize@url [0]{\catcode `\\12\catcode `\$12\catcode
  `\&12\catcode `\#12\catcode `\^12\catcode `\_12\catcode `\%12\relax}%
\providecommand \@@startlink[1]{}%
\providecommand \@@endlink[0]{}%
\providecommand \url  [0]{\begingroup\@sanitize@url \@url }%
\providecommand \@url [1]{\endgroup\@href {#1}{\urlprefix }}%
\providecommand \urlprefix  [0]{URL }%
\providecommand \Eprint [0]{\href }%
\@ifxundefined \urlstyle {%
  \providecommand \doi  [0]{\begingroup \@sanitize@url \@doi}%
  \providecommand \@doi [1]{\endgroup \@@startlink {\doibase
  #1}doi:\discretionary {}{}{}#1\@@endlink }%
}{%
  \providecommand \doi  [0]{doi:\discretionary{}{}{}\begingroup
  \urlstyle{rm}\Url }%
}%
\providecommand \doibase [0]{http://dx.doi.org/}%
\providecommand \Doi [0]{\begingroup \@sanitize@url \@Doi }%
\providecommand \@Doi  [1]{\endgroup\@@startlink{\doibase#1}\@@Doi}%
\providecommand \@@Doi [1]{#1\@@endlink}%
\providecommand \selectlanguage [0]{\@gobble}%
\providecommand \bibinfo  [0]{\@secondoftwo}%
\providecommand \bibfield  [0]{\@secondoftwo}%
\providecommand \translation [1]{[#1]}%
\providecommand \BibitemOpen [0]{}%
\providecommand \bibitemStop [0]{}%
\providecommand \bibitemNoStop [0]{.\EOS\space}%
\providecommand \EOS [0]{\spacefactor3000\relax}%
\providecommand \BibitemShut  [1]{\csname bibitem#1\endcsname}%
\bibitem [{\citenamefont {Okada}\ \emph {et~al.}(2016)\citenamefont {Okada},
  \citenamefont {Ogawa}, \citenamefont {Yoshimi}, \citenamefont {Tsukazaki},
  \citenamefont {Takahashi}, \citenamefont {Kawasaki},\ and\ \citenamefont
  {Tokura}}]{Okada16PRB_Fermi-level_TI}%
  \BibitemOpen
  \bibfield  {author} {\bibinfo {author} {\bibfnamefont {K.~N.}\ \bibnamefont
  {Okada}}, \bibinfo {author} {\bibfnamefont {N.}~\bibnamefont {Ogawa}},
  \bibinfo {author} {\bibfnamefont {R.}~\bibnamefont {Yoshimi}}, \bibinfo
  {author} {\bibfnamefont {A.}~\bibnamefont {Tsukazaki}}, \bibinfo {author}
  {\bibfnamefont {K.~S.}\ \bibnamefont {Takahashi}}, \bibinfo {author}
  {\bibfnamefont {M.}~\bibnamefont {Kawasaki}}, \ and\ \bibinfo {author}
  {\bibfnamefont {Y.}~\bibnamefont {Tokura}},\ }\href
  {https://link.aps.org/doi/10.1103/PhysRevB.93.081403} {\bibfield  {journal}
  {\bibinfo  {journal} {\emph {Phys. Rev. B}},\ }\textbf {\bibinfo {volume}
  {93}},\ \bibinfo {pages} {081403}\  (\bibinfo {year} {2016})}\BibitemShut
  {NoStop}%
\bibitem [{\citenamefont {Kondou}\ \emph {et~al.}(2016)\citenamefont {Kondou},
  \citenamefont {Yoshimi}, \citenamefont {Tsukazaki}, \citenamefont {Fukuma},
  \citenamefont {Matsuno}, \citenamefont {Takahashi}, \citenamefont {Kawasaki},
  \citenamefont {Tokura},\ and\ \citenamefont
  {Otani}}]{Kondou16NP_Fermi-level-sc_TI}%
  \BibitemOpen
  \bibfield  {author} {\bibinfo {author} {\bibfnamefont {K.}~\bibnamefont
  {Kondou}}, \bibinfo {author} {\bibfnamefont {R.}~\bibnamefont {Yoshimi}},
  \bibinfo {author} {\bibfnamefont {A.}~\bibnamefont {Tsukazaki}}, \bibinfo
  {author} {\bibfnamefont {Y.}~\bibnamefont {Fukuma}}, \bibinfo {author}
  {\bibfnamefont {J.}~\bibnamefont {Matsuno}}, \bibinfo {author} {\bibfnamefont
  {K.~S.}\ \bibnamefont {Takahashi}}, \bibinfo {author} {\bibfnamefont
  {M.}~\bibnamefont {Kawasaki}}, \bibinfo {author} {\bibfnamefont
  {Y.}~\bibnamefont {Tokura}}, \ and\ \bibinfo {author} {\bibfnamefont
  {Y.}~\bibnamefont {Otani}},\ }\href {https://doi.org/10.1038/nphys3833}
  {\bibfield  {journal} {\bibinfo  {journal} {\emph {Nat. Phys.}},\ }\textbf
  {\bibinfo {volume} {12}},\ \bibinfo {pages} {1027}\  (\bibinfo {year}
  {2016})}\BibitemShut {NoStop}%
\bibitem [{\citenamefont {Sun}\ \emph {et~al.}(2019)\citenamefont {Sun},
  \citenamefont {Yang}, \citenamefont {Yang}, \citenamefont {Vetter},
  \citenamefont {Sun}, \citenamefont {Li}, \citenamefont {Su}, \citenamefont
  {Li}, \citenamefont {Li}, \citenamefont {Gong} \emph
  {et~al.}}]{sun2019large}%
  \BibitemOpen
  \bibfield  {author} {\bibinfo {author} {\bibfnamefont {R.}~\bibnamefont
  {Sun}}, \bibinfo {author} {\bibfnamefont {S.}~\bibnamefont {Yang}}, \bibinfo
  {author} {\bibfnamefont {X.}~\bibnamefont {Yang}}, \bibinfo {author}
  {\bibfnamefont {E.}~\bibnamefont {Vetter}}, \bibinfo {author} {\bibfnamefont
  {D.}~\bibnamefont {Sun}}, \bibinfo {author} {\bibfnamefont {N.}~\bibnamefont
  {Li}}, \bibinfo {author} {\bibfnamefont {L.}~\bibnamefont {Su}}, \bibinfo
  {author} {\bibfnamefont {Y.}~\bibnamefont {Li}}, \bibinfo {author}
  {\bibfnamefont {Y.}~\bibnamefont {Li}}, \bibinfo {author} {\bibfnamefont
  {Z.-z.}\ \bibnamefont {Gong}},  \emph {et~al.},\ }\href
  {https://doi.org/10.1021/acs.nanolett.9b01151} {\bibfield  {journal}
  {\bibinfo  {journal} {\emph {Nano Lett.}},\ }\textbf {\bibinfo {volume}
  {19}},\ \bibinfo {pages} {4420}\  (\bibinfo {year} {2019})}\BibitemShut
  {NoStop}%
\bibitem [{\citenamefont {Su}\ \emph {et~al.}(2021)\citenamefont {Su},
  \citenamefont {Chuang}, \citenamefont {Lee}, \citenamefont {Chong},
  \citenamefont {Li}, \citenamefont {Lin}, \citenamefont {Chen}, \citenamefont
  {Cheng},\ and\ \citenamefont {Huang}}]{Su21ACS_Fermi-level_TI}%
  \BibitemOpen
  \bibfield  {author} {\bibinfo {author} {\bibfnamefont {S.~H.}\ \bibnamefont
  {Su}}, \bibinfo {author} {\bibfnamefont {P.-Y.}\ \bibnamefont {Chuang}},
  \bibinfo {author} {\bibfnamefont {J.-C.}\ \bibnamefont {Lee}}, \bibinfo
  {author} {\bibfnamefont {C.-W.}\ \bibnamefont {Chong}}, \bibinfo {author}
  {\bibfnamefont {Y.~W.}\ \bibnamefont {Li}}, \bibinfo {author} {\bibfnamefont
  {Z.~M.}\ \bibnamefont {Lin}}, \bibinfo {author} {\bibfnamefont {Y.-C.}\
  \bibnamefont {Chen}}, \bibinfo {author} {\bibfnamefont {C.-M.}\ \bibnamefont
  {Cheng}}, \ and\ \bibinfo {author} {\bibfnamefont {J.-C.-A.}\ \bibnamefont
  {Huang}},\ }\href {https://doi.org/10.1021/acsaelm.1c00182} {\bibfield
  {journal} {\bibinfo  {journal} {\emph {ACS Appl. Electron. Mater.}},\
  }\textbf {\bibinfo {volume} {3}},\ \bibinfo {pages} {2988}\  (\bibinfo {year}
  {2021})}\BibitemShut {NoStop}%
\bibitem [{\citenamefont {Qi}\ \emph {et~al.}(2006)\citenamefont {Qi},
  \citenamefont {Wu},\ and\ \citenamefont {Zhang}}]{qi2006topological}%
  \BibitemOpen
  \bibfield  {author} {\bibinfo {author} {\bibfnamefont {X.-L.}\ \bibnamefont
  {Qi}}, \bibinfo {author} {\bibfnamefont {Y.-S.}\ \bibnamefont {Wu}}, \ and\
  \bibinfo {author} {\bibfnamefont {S.-C.}\ \bibnamefont {Zhang}},\ }\href
  {https://link.aps.org/doi/10.1103/PhysRevB.74.085308} {\bibfield  {journal}
  {\bibinfo  {journal} {\emph {Phys. Rev. B}},\ }\textbf {\bibinfo {volume}
  {74}},\ \bibinfo {pages} {085308}\  (\bibinfo {year} {2006})}\BibitemShut
  {NoStop}%
\bibitem [{\citenamefont {Yu}\ \emph {et~al.}(2010)\citenamefont {Yu},
  \citenamefont {Zhang}, \citenamefont {Zhang}, \citenamefont {Zhang},
  \citenamefont {Dai},\ and\ \citenamefont {Fang}}]{yu2010quantized}%
  \BibitemOpen
  \bibfield  {author} {\bibinfo {author} {\bibfnamefont {R.}~\bibnamefont
  {Yu}}, \bibinfo {author} {\bibfnamefont {W.}~\bibnamefont {Zhang}}, \bibinfo
  {author} {\bibfnamefont {H.-J.}\ \bibnamefont {Zhang}}, \bibinfo {author}
  {\bibfnamefont {S.-C.}\ \bibnamefont {Zhang}}, \bibinfo {author}
  {\bibfnamefont {X.}~\bibnamefont {Dai}}, \ and\ \bibinfo {author}
  {\bibfnamefont {Z.}~\bibnamefont {Fang}},\ }\href
  {https://doi.org/10.1126/science.1187485} {\bibfield  {journal} {\bibinfo
  {journal} {\emph {Science}},\ }\textbf {\bibinfo {volume} {329}},\ \bibinfo
  {pages} {61}\  (\bibinfo {year} {2010})}\BibitemShut {NoStop}%
\bibitem [{\citenamefont {Chang}\ \emph {et~al.}(2013)\citenamefont {Chang},
  \citenamefont {Zhang}, \citenamefont {Feng}, \citenamefont {Shen},
  \citenamefont {Zhang}, \citenamefont {Guo}, \citenamefont {Li}, \citenamefont
  {Ou}, \citenamefont {Wei}, \citenamefont {Wang} \emph
  {et~al.}}]{chang2013experimental}%
  \BibitemOpen
  \bibfield  {author} {\bibinfo {author} {\bibfnamefont {C.-Z.}\ \bibnamefont
  {Chang}}, \bibinfo {author} {\bibfnamefont {J.}~\bibnamefont {Zhang}},
  \bibinfo {author} {\bibfnamefont {X.}~\bibnamefont {Feng}}, \bibinfo {author}
  {\bibfnamefont {J.}~\bibnamefont {Shen}}, \bibinfo {author} {\bibfnamefont
  {Z.}~\bibnamefont {Zhang}}, \bibinfo {author} {\bibfnamefont
  {M.}~\bibnamefont {Guo}}, \bibinfo {author} {\bibfnamefont {K.}~\bibnamefont
  {Li}}, \bibinfo {author} {\bibfnamefont {Y.}~\bibnamefont {Ou}}, \bibinfo
  {author} {\bibfnamefont {P.}~\bibnamefont {Wei}}, \bibinfo {author}
  {\bibfnamefont {L.-L.}\ \bibnamefont {Wang}},  \emph {et~al.},\ }\href
  {https://doi.org/10.1126/science.1234414} {\bibfield  {journal} {\bibinfo
  {journal} {\emph {Science}},\ }\textbf {\bibinfo {volume} {340}},\ \bibinfo
  {pages} {167}\  (\bibinfo {year} {2013})}\BibitemShut {NoStop}%
\bibitem [{\citenamefont {Checkelsky}\ \emph {et~al.}(2014)\citenamefont
  {Checkelsky}, \citenamefont {Yoshimi}, \citenamefont {Tsukazaki},
  \citenamefont {Takahashi}, \citenamefont {Kozuka}, \citenamefont {Falson},
  \citenamefont {Kawasaki},\ and\ \citenamefont
  {Tokura}}]{checkelsky2014trajectory}%
  \BibitemOpen
  \bibfield  {author} {\bibinfo {author} {\bibfnamefont {J.}~\bibnamefont
  {Checkelsky}}, \bibinfo {author} {\bibfnamefont {R.}~\bibnamefont {Yoshimi}},
  \bibinfo {author} {\bibfnamefont {A.}~\bibnamefont {Tsukazaki}}, \bibinfo
  {author} {\bibfnamefont {K.}~\bibnamefont {Takahashi}}, \bibinfo {author}
  {\bibfnamefont {Y.}~\bibnamefont {Kozuka}}, \bibinfo {author} {\bibfnamefont
  {J.}~\bibnamefont {Falson}}, \bibinfo {author} {\bibfnamefont
  {M.}~\bibnamefont {Kawasaki}}, \ and\ \bibinfo {author} {\bibfnamefont
  {Y.}~\bibnamefont {Tokura}},\ }\href {https://doi.org/10.1038/nphys3053}
  {\bibfield  {journal} {\bibinfo  {journal} {\emph {Nat. Phys.}},\ }\textbf
  {\bibinfo {volume} {10}},\ \bibinfo {pages} {731}\  (\bibinfo {year}
  {2014})}\BibitemShut {NoStop}%
\bibitem [{\citenamefont {Liu}\ \emph {et~al.}(2016)\citenamefont {Liu},
  \citenamefont {Zhang},\ and\ \citenamefont {Qi}}]{liu2016quantum}%
  \BibitemOpen
  \bibfield  {author} {\bibinfo {author} {\bibfnamefont {C.-X.}\ \bibnamefont
  {Liu}}, \bibinfo {author} {\bibfnamefont {S.-C.}\ \bibnamefont {Zhang}}, \
  and\ \bibinfo {author} {\bibfnamefont {X.-L.}\ \bibnamefont {Qi}},\ }\href
  {https://doi.org/10.1146/annurev-conmatphys-031115-011417} {\bibfield
  {journal} {\bibinfo  {journal} {\emph {Annu. Rev. Condens. Matter Phys.}},\
  }\textbf {\bibinfo {volume} {7}},\ \bibinfo {pages} {301}\  (\bibinfo {year}
  {2016})}\BibitemShut {NoStop}%
\bibitem [{\citenamefont {Chang}\ \emph {et~al.}(2023)\citenamefont {Chang},
  \citenamefont {Liu},\ and\ \citenamefont {MacDonald}}]{chang2023colloquium}%
  \BibitemOpen
  \bibfield  {author} {\bibinfo {author} {\bibfnamefont {C.-Z.}\ \bibnamefont
  {Chang}}, \bibinfo {author} {\bibfnamefont {C.-X.}\ \bibnamefont {Liu}}, \
  and\ \bibinfo {author} {\bibfnamefont {A.~H.}\ \bibnamefont {MacDonald}},\
  }\href {https://link.aps.org/doi/10.1103/RevModPhys.95.011002} {\bibfield
  {journal} {\bibinfo  {journal} {\emph {Rev. Mod. Phys.}},\ }\textbf {\bibinfo
  {volume} {95}},\ \bibinfo {pages} {011002}\  (\bibinfo {year}
  {2023})}\BibitemShut {NoStop}%
\bibitem [{\citenamefont {Wu}\ \emph {et~al.}(2020)\citenamefont {Wu},
  \citenamefont {Gro{\ss}}, \citenamefont {Dai}, \citenamefont {Lujan},
  \citenamefont {Razavi}, \citenamefont {Zhang}, \citenamefont {Liu},
  \citenamefont {Sobotkiewich}, \citenamefont {F{\"o}rster}, \citenamefont
  {Weigand} \emph {et~al.}}]{wu2020ferrimagnetic}%
  \BibitemOpen
  \bibfield  {author} {\bibinfo {author} {\bibfnamefont {H.}~\bibnamefont
  {Wu}}, \bibinfo {author} {\bibfnamefont {F.}~\bibnamefont {Gro{\ss}}},
  \bibinfo {author} {\bibfnamefont {B.}~\bibnamefont {Dai}}, \bibinfo {author}
  {\bibfnamefont {D.}~\bibnamefont {Lujan}}, \bibinfo {author} {\bibfnamefont
  {S.~A.}\ \bibnamefont {Razavi}}, \bibinfo {author} {\bibfnamefont
  {P.}~\bibnamefont {Zhang}}, \bibinfo {author} {\bibfnamefont
  {Y.}~\bibnamefont {Liu}}, \bibinfo {author} {\bibfnamefont {K.}~\bibnamefont
  {Sobotkiewich}}, \bibinfo {author} {\bibfnamefont {J.}~\bibnamefont
  {F{\"o}rster}}, \bibinfo {author} {\bibfnamefont {M.}~\bibnamefont
  {Weigand}},  \emph {et~al.},\ }\href
  {https://onlinelibrary.wiley.com/doi/abs/10.1002/adma.202003380} {\bibfield
  {journal} {\bibinfo  {journal} {\emph {Adv. Mater.}},\ }\textbf {\bibinfo
  {volume} {32}},\ \bibinfo {pages} {2003380}\  (\bibinfo {year}
  {2020})}\BibitemShut {NoStop}%
\bibitem [{\citenamefont {Li}\ \emph {et~al.}(2021)\citenamefont {Li},
  \citenamefont {Ding}, \citenamefont {Zhang}, \citenamefont {Kally},
  \citenamefont {Pillsbury}, \citenamefont {Heinonen}, \citenamefont {Rimal},
  \citenamefont {Bi}, \citenamefont {DeMann}, \citenamefont {Field} \emph
  {et~al.}}]{li2021topological}%
  \BibitemOpen
  \bibfield  {author} {\bibinfo {author} {\bibfnamefont {P.}~\bibnamefont
  {Li}}, \bibinfo {author} {\bibfnamefont {J.}~\bibnamefont {Ding}}, \bibinfo
  {author} {\bibfnamefont {S.~S.-L.}\ \bibnamefont {Zhang}}, \bibinfo {author}
  {\bibfnamefont {J.}~\bibnamefont {Kally}}, \bibinfo {author} {\bibfnamefont
  {T.}~\bibnamefont {Pillsbury}}, \bibinfo {author} {\bibfnamefont {O.~G.}\
  \bibnamefont {Heinonen}}, \bibinfo {author} {\bibfnamefont {G.}~\bibnamefont
  {Rimal}}, \bibinfo {author} {\bibfnamefont {C.}~\bibnamefont {Bi}}, \bibinfo
  {author} {\bibfnamefont {A.}~\bibnamefont {DeMann}}, \bibinfo {author}
  {\bibfnamefont {S.~B.}\ \bibnamefont {Field}},  \emph {et~al.},\ }\href
  {https://doi.org/10.1021/acs.nanolett.0c03195} {\bibfield  {journal}
  {\bibinfo  {journal} {\emph {Nano Lett.}},\ }\textbf {\bibinfo {volume}
  {21}},\ \bibinfo {pages} {84}\  (\bibinfo {year} {2021})}\BibitemShut
  {NoStop}%
\bibitem [{\citenamefont {Zhang}\ \emph {et~al.}(2021)\citenamefont {Zhang},
  \citenamefont {Ambhire}, \citenamefont {Lu}, \citenamefont {Niu},
  \citenamefont {Cook}, \citenamefont {Jiang}, \citenamefont {Hong},
  \citenamefont {Alahmed}, \citenamefont {He}, \citenamefont {Zhang} \emph
  {et~al.}}]{zhang2021giant}%
  \BibitemOpen
  \bibfield  {author} {\bibinfo {author} {\bibfnamefont {X.}~\bibnamefont
  {Zhang}}, \bibinfo {author} {\bibfnamefont {S.~C.}\ \bibnamefont {Ambhire}},
  \bibinfo {author} {\bibfnamefont {Q.}~\bibnamefont {Lu}}, \bibinfo {author}
  {\bibfnamefont {W.}~\bibnamefont {Niu}}, \bibinfo {author} {\bibfnamefont
  {J.}~\bibnamefont {Cook}}, \bibinfo {author} {\bibfnamefont {J.~S.}\
  \bibnamefont {Jiang}}, \bibinfo {author} {\bibfnamefont {D.}~\bibnamefont
  {Hong}}, \bibinfo {author} {\bibfnamefont {L.}~\bibnamefont {Alahmed}},
  \bibinfo {author} {\bibfnamefont {L.}~\bibnamefont {He}}, \bibinfo {author}
  {\bibfnamefont {R.}~\bibnamefont {Zhang}},  \emph {et~al.},\ }\href
  {https://doi.org/10.1021/acsnano.1c05519} {\bibfield  {journal} {\bibinfo
  {journal} {\emph {ACS Nano}},\ }\textbf {\bibinfo {volume} {15}},\ \bibinfo
  {pages} {15710}\  (\bibinfo {year} {2021})}\BibitemShut {NoStop}%
\bibitem [{\citenamefont {Mellnik}\ \emph {et~al.}(2014)\citenamefont
  {Mellnik}, \citenamefont {Lee}, \citenamefont {Richardella}, \citenamefont
  {Grab}, \citenamefont {Mintun}, \citenamefont {Fischer}, \citenamefont
  {Vaezi}, \citenamefont {Manchon}, \citenamefont {Kim}, \citenamefont
  {Samarth} \emph {et~al.}}]{Mellnik2014}%
  \BibitemOpen
  \bibfield  {author} {\bibinfo {author} {\bibfnamefont {A.~R.}\ \bibnamefont
  {Mellnik}}, \bibinfo {author} {\bibfnamefont {J.~S.}\ \bibnamefont {Lee}},
  \bibinfo {author} {\bibfnamefont {A.}~\bibnamefont {Richardella}}, \bibinfo
  {author} {\bibfnamefont {J.~L.}\ \bibnamefont {Grab}}, \bibinfo {author}
  {\bibfnamefont {P.~J.}\ \bibnamefont {Mintun}}, \bibinfo {author}
  {\bibfnamefont {M.~H.}\ \bibnamefont {Fischer}}, \bibinfo {author}
  {\bibfnamefont {A.}~\bibnamefont {Vaezi}}, \bibinfo {author} {\bibfnamefont
  {A.}~\bibnamefont {Manchon}}, \bibinfo {author} {\bibfnamefont {E.-A.}\
  \bibnamefont {Kim}}, \bibinfo {author} {\bibfnamefont {N.}~\bibnamefont
  {Samarth}},  \emph {et~al.},\ }\href {https://doi.org/10.1038/nature13534}
  {\bibfield  {journal} {\bibinfo  {journal} {\emph {Nature}},\ }\textbf
  {\bibinfo {volume} {511}},\ \bibinfo {pages} {449}\  (\bibinfo {year}
  {2014})}\BibitemShut {NoStop}%
\bibitem [{\citenamefont {Han}\ \emph {et~al.}(2017)\citenamefont {Han},
  \citenamefont {Richardella}, \citenamefont {Siddiqui}, \citenamefont
  {Finley}, \citenamefont {Samarth},\ and\ \citenamefont
  {Liu}}]{Han17PRL_SOT-TI}%
  \BibitemOpen
  \bibfield  {author} {\bibinfo {author} {\bibfnamefont {J.}~\bibnamefont
  {Han}}, \bibinfo {author} {\bibfnamefont {A.}~\bibnamefont {Richardella}},
  \bibinfo {author} {\bibfnamefont {S.~A.}\ \bibnamefont {Siddiqui}}, \bibinfo
  {author} {\bibfnamefont {J.}~\bibnamefont {Finley}}, \bibinfo {author}
  {\bibfnamefont {N.}~\bibnamefont {Samarth}}, \ and\ \bibinfo {author}
  {\bibfnamefont {L.}~\bibnamefont {Liu}},\ }\href
  {https://link.aps.org/doi/10.1103/PhysRevLett.119.077702} {\bibfield
  {journal} {\bibinfo  {journal} {\emph {Phys. Rev. Lett.}},\ }\textbf
  {\bibinfo {volume} {119}},\ \bibinfo {pages} {077702}\  (\bibinfo {year}
  {2017})}\BibitemShut {NoStop}%
\bibitem [{\citenamefont {Li}\ \emph {et~al.}(2019)\citenamefont {Li},
  \citenamefont {Kally}, \citenamefont {Zhang}, \citenamefont {Pillsbury},
  \citenamefont {Ding}, \citenamefont {Csaba}, \citenamefont {Ding},
  \citenamefont {Jiang}, \citenamefont {Liu}, \citenamefont {Sinclair} \emph
  {et~al.}}]{li2019magnetization}%
  \BibitemOpen
  \bibfield  {author} {\bibinfo {author} {\bibfnamefont {P.}~\bibnamefont
  {Li}}, \bibinfo {author} {\bibfnamefont {J.}~\bibnamefont {Kally}}, \bibinfo
  {author} {\bibfnamefont {S.~S.-L.}\ \bibnamefont {Zhang}}, \bibinfo {author}
  {\bibfnamefont {T.}~\bibnamefont {Pillsbury}}, \bibinfo {author}
  {\bibfnamefont {J.}~\bibnamefont {Ding}}, \bibinfo {author} {\bibfnamefont
  {G.}~\bibnamefont {Csaba}}, \bibinfo {author} {\bibfnamefont
  {J.}~\bibnamefont {Ding}}, \bibinfo {author} {\bibfnamefont {J.}~\bibnamefont
  {Jiang}}, \bibinfo {author} {\bibfnamefont {Y.}~\bibnamefont {Liu}}, \bibinfo
  {author} {\bibfnamefont {R.}~\bibnamefont {Sinclair}},  \emph {et~al.},\
  }\href {https://doi.org/10.1126/sciadv.aaw3415} {\bibfield  {journal}
  {\bibinfo  {journal} {\emph {Sci. Adv.}},\ }\textbf {\bibinfo {volume} {5}},\
  \bibinfo {pages} {eaaw3415}\  (\bibinfo {year} {2019})}\BibitemShut {NoStop}%
\bibitem [{\citenamefont {Moghaddam}\ \emph {et~al.}(2020)\citenamefont
  {Moghaddam}, \citenamefont {Qaiumzadeh}, \citenamefont {Dyrda\l{}},\ and\
  \citenamefont {Berakdar}}]{Moghaddam20PRL_SOT-AMR-proximity-TI}%
  \BibitemOpen
  \bibfield  {author} {\bibinfo {author} {\bibfnamefont {A.~G.}\ \bibnamefont
  {Moghaddam}}, \bibinfo {author} {\bibfnamefont {A.}~\bibnamefont
  {Qaiumzadeh}}, \bibinfo {author} {\bibfnamefont {A.}~\bibnamefont
  {Dyrda\l{}}}, \ and\ \bibinfo {author} {\bibfnamefont {J.}~\bibnamefont
  {Berakdar}},\ }\href
  {https://link.aps.org/doi/10.1103/PhysRevLett.125.196801} {\bibfield
  {journal} {\bibinfo  {journal} {\emph {Phys. Rev. Lett.}},\ }\textbf
  {\bibinfo {volume} {125}},\ \bibinfo {pages} {196801}\  (\bibinfo {year}
  {2020})}\BibitemShut {NoStop}%
\bibitem [{\citenamefont {Chiba}\ \emph {et~al.}(2017)\citenamefont {Chiba},
  \citenamefont {Takahashi},\ and\ \citenamefont {Bauer}}]{chiba2017magnetic}%
  \BibitemOpen
  \bibfield  {author} {\bibinfo {author} {\bibfnamefont {T.}~\bibnamefont
  {Chiba}}, \bibinfo {author} {\bibfnamefont {S.}~\bibnamefont {Takahashi}}, \
  and\ \bibinfo {author} {\bibfnamefont {G.~E.~W.}\ \bibnamefont {Bauer}},\
  }\href {https://link.aps.org/doi/10.1103/PhysRevB.95.094428} {\bibfield
  {journal} {\bibinfo  {journal} {\emph {Phys. Rev. B}},\ }\textbf {\bibinfo
  {volume} {95}},\ \bibinfo {pages} {094428}\  (\bibinfo {year}
  {2017})}\BibitemShut {NoStop}%
\bibitem [{\citenamefont {Sklenar}\ \emph {et~al.}(2021)\citenamefont
  {Sklenar}, \citenamefont {Zhang}, \citenamefont {Jungfleisch}, \citenamefont
  {Kim}, \citenamefont {Xiao}, \citenamefont {MacDougall}, \citenamefont
  {Gilbert}, \citenamefont {Hoffmann}, \citenamefont {Schiffer},\ and\
  \citenamefont {Mason}}]{sklenar2021proximity}%
  \BibitemOpen
  \bibfield  {author} {\bibinfo {author} {\bibfnamefont {J.}~\bibnamefont
  {Sklenar}}, \bibinfo {author} {\bibfnamefont {Y.}~\bibnamefont {Zhang}},
  \bibinfo {author} {\bibfnamefont {M.~B.}\ \bibnamefont {Jungfleisch}},
  \bibinfo {author} {\bibfnamefont {Y.}~\bibnamefont {Kim}}, \bibinfo {author}
  {\bibfnamefont {Y.}~\bibnamefont {Xiao}}, \bibinfo {author} {\bibfnamefont
  {G.~J.}\ \bibnamefont {MacDougall}}, \bibinfo {author} {\bibfnamefont
  {M.~J.}\ \bibnamefont {Gilbert}}, \bibinfo {author} {\bibfnamefont
  {A.}~\bibnamefont {Hoffmann}}, \bibinfo {author} {\bibfnamefont
  {P.}~\bibnamefont {Schiffer}}, \ and\ \bibinfo {author} {\bibfnamefont
  {N.}~\bibnamefont {Mason}},\ }\href {https://doi.org/10.1063/5.0052301}
  {\bibfield  {journal} {\bibinfo  {journal} {\emph {Appl. Phys. Lett.}},\
  }\textbf {\bibinfo {volume} {118}},\ \bibinfo {pages} {232402}\  (\bibinfo
  {year} {2021})}\BibitemShut {NoStop}%
\bibitem [{\citenamefont {Wu}\ \emph {et~al.}(2021)\citenamefont {Wu},
  \citenamefont {Chen}, \citenamefont {Zhang}, \citenamefont {He},
  \citenamefont {Nance}, \citenamefont {Guo}, \citenamefont {Sasaki},
  \citenamefont {Shirokura}, \citenamefont {Hai}, \citenamefont {Fang} \emph
  {et~al.}}]{Wu21NC_MRAM-TI}%
  \BibitemOpen
  \bibfield  {author} {\bibinfo {author} {\bibfnamefont {H.}~\bibnamefont
  {Wu}}, \bibinfo {author} {\bibfnamefont {A.}~\bibnamefont {Chen}}, \bibinfo
  {author} {\bibfnamefont {P.}~\bibnamefont {Zhang}}, \bibinfo {author}
  {\bibfnamefont {H.}~\bibnamefont {He}}, \bibinfo {author} {\bibfnamefont
  {J.}~\bibnamefont {Nance}}, \bibinfo {author} {\bibfnamefont
  {C.}~\bibnamefont {Guo}}, \bibinfo {author} {\bibfnamefont {J.}~\bibnamefont
  {Sasaki}}, \bibinfo {author} {\bibfnamefont {T.}~\bibnamefont {Shirokura}},
  \bibinfo {author} {\bibfnamefont {P.~N.}\ \bibnamefont {Hai}}, \bibinfo
  {author} {\bibfnamefont {B.}~\bibnamefont {Fang}},  \emph {et~al.},\ }\href
  {https://doi.org/10.1038/s41467-021-26478-3} {\bibfield  {journal} {\bibinfo
  {journal} {\emph {Nat. Commun.}},\ }\textbf {\bibinfo {volume} {12}},\
  \bibinfo {pages} {6251}\  (\bibinfo {year} {2021})}\BibitemShut {NoStop}%
\bibitem [{\citenamefont {Fu}\ and\ \citenamefont
  {Kane}(2008)}]{Fu08PRL_proximity-SC_TI_majorana}%
  \BibitemOpen
  \bibfield  {author} {\bibinfo {author} {\bibfnamefont {L.}~\bibnamefont
  {Fu}}\ and\ \bibinfo {author} {\bibfnamefont {C.~L.}\ \bibnamefont {Kane}},\
  }\href {https://link.aps.org/doi/10.1103/PhysRevLett.100.096407} {\bibfield
  {journal} {\bibinfo  {journal} {\emph {Phys. Rev. Lett.}},\ }\textbf
  {\bibinfo {volume} {100}},\ \bibinfo {pages} {096407}\  (\bibinfo {year}
  {2008})}\BibitemShut {NoStop}%
\bibitem [{\citenamefont {Lian}\ \emph {et~al.}(2018)\citenamefont {Lian},
  \citenamefont {Sun}, \citenamefont {Vaezi}, \citenamefont {Qi},\ and\
  \citenamefont {Zhang}}]{Lian18PNAS_Topo-quant-comput_MZM}%
  \BibitemOpen
  \bibfield  {author} {\bibinfo {author} {\bibfnamefont {B.}~\bibnamefont
  {Lian}}, \bibinfo {author} {\bibfnamefont {X.-Q.}\ \bibnamefont {Sun}},
  \bibinfo {author} {\bibfnamefont {A.}~\bibnamefont {Vaezi}}, \bibinfo
  {author} {\bibfnamefont {X.-L.}\ \bibnamefont {Qi}}, \ and\ \bibinfo {author}
  {\bibfnamefont {S.-C.}\ \bibnamefont {Zhang}},\ }\href
  {https://www.pnas.org/doi/abs/10.1073/pnas.1810003115} {\bibfield  {journal}
  {\bibinfo  {journal} {\emph {Proc. Natl. Acad. Sci.}},\ }\textbf {\bibinfo
  {volume} {115}},\ \bibinfo {pages} {10938}\  (\bibinfo {year}
  {2018})}\BibitemShut {NoStop}%
\bibitem [{\citenamefont {Yasuda}\ \emph {et~al.}(2016)\citenamefont {Yasuda},
  \citenamefont {Tsukazaki}, \citenamefont {Yoshimi}, \citenamefont
  {Takahashi}, \citenamefont {Kawasaki},\ and\ \citenamefont
  {Tokura}}]{yasuda2016large}%
  \BibitemOpen
  \bibfield  {author} {\bibinfo {author} {\bibfnamefont {K.}~\bibnamefont
  {Yasuda}}, \bibinfo {author} {\bibfnamefont {A.}~\bibnamefont {Tsukazaki}},
  \bibinfo {author} {\bibfnamefont {R.}~\bibnamefont {Yoshimi}}, \bibinfo
  {author} {\bibfnamefont {K.~S.}\ \bibnamefont {Takahashi}}, \bibinfo {author}
  {\bibfnamefont {M.}~\bibnamefont {Kawasaki}}, \ and\ \bibinfo {author}
  {\bibfnamefont {Y.}~\bibnamefont {Tokura}},\ }\href
  {https://link.aps.org/doi/10.1103/PhysRevLett.117.127202} {\bibfield
  {journal} {\bibinfo  {journal} {\emph {Phys. Rev. Lett.}},\ }\textbf
  {\bibinfo {volume} {117}},\ \bibinfo {pages} {127202}\  (\bibinfo {year}
  {2016})}\BibitemShut {NoStop}%
\bibitem [{\citenamefont {Yasuda}\ \emph {et~al.}(2017)\citenamefont {Yasuda},
  \citenamefont {Tsukazaki}, \citenamefont {Yoshimi}, \citenamefont {Kondou},
  \citenamefont {Takahashi}, \citenamefont {Otani}, \citenamefont {Kawasaki},\
  and\ \citenamefont {Tokura}}]{yasuda2017current}%
  \BibitemOpen
  \bibfield  {author} {\bibinfo {author} {\bibfnamefont {K.}~\bibnamefont
  {Yasuda}}, \bibinfo {author} {\bibfnamefont {A.}~\bibnamefont {Tsukazaki}},
  \bibinfo {author} {\bibfnamefont {R.}~\bibnamefont {Yoshimi}}, \bibinfo
  {author} {\bibfnamefont {K.}~\bibnamefont {Kondou}}, \bibinfo {author}
  {\bibfnamefont {K.~S.}\ \bibnamefont {Takahashi}}, \bibinfo {author}
  {\bibfnamefont {Y.}~\bibnamefont {Otani}}, \bibinfo {author} {\bibfnamefont
  {M.}~\bibnamefont {Kawasaki}}, \ and\ \bibinfo {author} {\bibfnamefont
  {Y.}~\bibnamefont {Tokura}},\ }\href
  {https://link.aps.org/doi/10.1103/PhysRevLett.119.137204} {\bibfield
  {journal} {\bibinfo  {journal} {\emph {Phys. Rev. Lett.}},\ }\textbf
  {\bibinfo {volume} {119}},\ \bibinfo {pages} {137204}\  (\bibinfo {year}
  {2017})}\BibitemShut {NoStop}%
\bibitem [{\citenamefont {Lv}\ \emph {et~al.}(2018)\citenamefont {Lv},
  \citenamefont {Kally}, \citenamefont {Zhang}, \citenamefont {Lee},
  \citenamefont {Jamali}, \citenamefont {Samarth},\ and\ \citenamefont
  {Wang}}]{lv2018unidirectional}%
  \BibitemOpen
  \bibfield  {author} {\bibinfo {author} {\bibfnamefont {Y.}~\bibnamefont
  {Lv}}, \bibinfo {author} {\bibfnamefont {J.}~\bibnamefont {Kally}}, \bibinfo
  {author} {\bibfnamefont {D.}~\bibnamefont {Zhang}}, \bibinfo {author}
  {\bibfnamefont {J.~S.}\ \bibnamefont {Lee}}, \bibinfo {author} {\bibfnamefont
  {M.}~\bibnamefont {Jamali}}, \bibinfo {author} {\bibfnamefont
  {N.}~\bibnamefont {Samarth}}, \ and\ \bibinfo {author} {\bibfnamefont
  {J.-P.}\ \bibnamefont {Wang}},\ }\href
  {https://doi.org/10.1038/s41467-017-02491-3} {\bibfield  {journal} {\bibinfo
  {journal} {\emph {Nat. Commun.}},\ }\textbf {\bibinfo {volume} {9}},\
  \bibinfo {pages} {1}\  (\bibinfo {year} {2018})}\BibitemShut {NoStop}%
\bibitem [{\citenamefont {Duy~Khang}\ and\ \citenamefont
  {Hai}(2019)}]{duy2019giant}%
  \BibitemOpen
  \bibfield  {author} {\bibinfo {author} {\bibfnamefont {N.~H.}\ \bibnamefont
  {Duy~Khang}}\ and\ \bibinfo {author} {\bibfnamefont {P.~N.}\ \bibnamefont
  {Hai}},\ }\href {https://doi.org/10.1063/1.5134728} {\bibfield  {journal}
  {\bibinfo  {journal} {\emph {J. Appl. Phys.}},\ }\textbf {\bibinfo {volume}
  {126}}\  (\bibinfo {year} {2019})}\BibitemShut {NoStop}%
\bibitem [{\citenamefont {Wang}\ \emph {et~al.}(2022)\citenamefont {Wang},
  \citenamefont {Mambakkam}, \citenamefont {Huang}, \citenamefont {Wang},
  \citenamefont {Ji}, \citenamefont {Xiao}, \citenamefont {Yang}, \citenamefont
  {Law},\ and\ \citenamefont {Xiao}}]{wang2022observation}%
  \BibitemOpen
  \bibfield  {author} {\bibinfo {author} {\bibfnamefont {Y.}~\bibnamefont
  {Wang}}, \bibinfo {author} {\bibfnamefont {S.~V.}\ \bibnamefont {Mambakkam}},
  \bibinfo {author} {\bibfnamefont {Y.-X.}\ \bibnamefont {Huang}}, \bibinfo
  {author} {\bibfnamefont {Y.}~\bibnamefont {Wang}}, \bibinfo {author}
  {\bibfnamefont {Y.}~\bibnamefont {Ji}}, \bibinfo {author} {\bibfnamefont
  {C.}~\bibnamefont {Xiao}}, \bibinfo {author} {\bibfnamefont {S.~A.}\
  \bibnamefont {Yang}}, \bibinfo {author} {\bibfnamefont {S.~A.}\ \bibnamefont
  {Law}}, \ and\ \bibinfo {author} {\bibfnamefont {J.~Q.}\ \bibnamefont
  {Xiao}},\ }\href {https://link.aps.org/doi/10.1103/PhysRevB.106.155408}
  {\bibfield  {journal} {\bibinfo  {journal} {\emph {Phys. Rev. B}},\ }\textbf
  {\bibinfo {volume} {106}},\ \bibinfo {pages} {155408}\  (\bibinfo {year}
  {2022})}\BibitemShut {NoStop}%
\bibitem [{\citenamefont {Lv}\ \emph {et~al.}(2022)\citenamefont {Lv},
  \citenamefont {Kally}, \citenamefont {Liu}, \citenamefont {Quarterman},
  \citenamefont {Pillsbury}, \citenamefont {Kirby}, \citenamefont {Grutter},
  \citenamefont {Sahu}, \citenamefont {Borchers}, \citenamefont {Wu} \emph
  {et~al.}}]{lv2022large}%
  \BibitemOpen
  \bibfield  {author} {\bibinfo {author} {\bibfnamefont {Y.}~\bibnamefont
  {Lv}}, \bibinfo {author} {\bibfnamefont {J.}~\bibnamefont {Kally}}, \bibinfo
  {author} {\bibfnamefont {T.}~\bibnamefont {Liu}}, \bibinfo {author}
  {\bibfnamefont {P.}~\bibnamefont {Quarterman}}, \bibinfo {author}
  {\bibfnamefont {T.}~\bibnamefont {Pillsbury}}, \bibinfo {author}
  {\bibfnamefont {B.~J.}\ \bibnamefont {Kirby}}, \bibinfo {author}
  {\bibfnamefont {A.~J.}\ \bibnamefont {Grutter}}, \bibinfo {author}
  {\bibfnamefont {P.}~\bibnamefont {Sahu}}, \bibinfo {author} {\bibfnamefont
  {J.~A.}\ \bibnamefont {Borchers}}, \bibinfo {author} {\bibfnamefont
  {M.}~\bibnamefont {Wu}},  \emph {et~al.},\ }\href
  {https://doi.org/10.1063/5.0073976} {\bibfield  {journal} {\bibinfo
  {journal} {\emph {Appl. Phys. Rev.}},\ }\textbf {\bibinfo {volume} {9}},\
  \bibinfo {pages} {011406}\  (\bibinfo {year} {2022})}\BibitemShut {NoStop}%
\bibitem [{\citenamefont {Avci}\ \emph {et~al.}(2015)\citenamefont {Avci},
  \citenamefont {Garello}, \citenamefont {Ghosh}, \citenamefont {Gabureac},
  \citenamefont {Alvarado},\ and\ \citenamefont
  {Gambardella}}]{avci2015unidirectional}%
  \BibitemOpen
  \bibfield  {author} {\bibinfo {author} {\bibfnamefont {C.~O.}\ \bibnamefont
  {Avci}}, \bibinfo {author} {\bibfnamefont {K.}~\bibnamefont {Garello}},
  \bibinfo {author} {\bibfnamefont {A.}~\bibnamefont {Ghosh}}, \bibinfo
  {author} {\bibfnamefont {M.}~\bibnamefont {Gabureac}}, \bibinfo {author}
  {\bibfnamefont {S.~F.}\ \bibnamefont {Alvarado}}, \ and\ \bibinfo {author}
  {\bibfnamefont {P.}~\bibnamefont {Gambardella}},\ }\href
  {https://doi.org/10.1038/nphys3356} {\bibfield  {journal} {\bibinfo
  {journal} {\emph {Nat. Phys.}},\ }\textbf {\bibinfo {volume} {11}},\ \bibinfo
  {pages} {570}\  (\bibinfo {year} {2015})}\BibitemShut {NoStop}%
\bibitem [{\citenamefont {Olejn\'{\i}k}\ \emph {et~al.}(2015)\citenamefont
  {Olejn\'{\i}k}, \citenamefont {Nov\'ak}, \citenamefont {Wunderlich},\ and\
  \citenamefont {Jungwirth}}]{olejnik2015electrical}%
  \BibitemOpen
  \bibfield  {author} {\bibinfo {author} {\bibfnamefont {K.}~\bibnamefont
  {Olejn\'{\i}k}}, \bibinfo {author} {\bibfnamefont {V.}~\bibnamefont
  {Nov\'ak}}, \bibinfo {author} {\bibfnamefont {J.}~\bibnamefont {Wunderlich}},
  \ and\ \bibinfo {author} {\bibfnamefont {T.}~\bibnamefont {Jungwirth}},\
  }\href {https://link.aps.org/doi/10.1103/PhysRevB.91.180402} {\bibfield
  {journal} {\bibinfo  {journal} {\emph {Phys. Rev. B}},\ }\textbf {\bibinfo
  {volume} {91}},\ \bibinfo {pages} {180402}\  (\bibinfo {year}
  {2015})}\BibitemShut {NoStop}%
\bibitem [{\citenamefont {Zhang}\ and\ \citenamefont
  {Vignale}(2016)}]{zhang2016theory}%
  \BibitemOpen
  \bibfield  {author} {\bibinfo {author} {\bibfnamefont {S.~S.-L.}\
  \bibnamefont {Zhang}}\ and\ \bibinfo {author} {\bibfnamefont
  {G.}~\bibnamefont {Vignale}},\ }\href
  {https://link.aps.org/doi/10.1103/PhysRevB.94.140411} {\bibfield  {journal}
  {\bibinfo  {journal} {\emph {Phys. Rev. B}},\ }\textbf {\bibinfo {volume}
  {94}},\ \bibinfo {pages} {140411}\  (\bibinfo {year} {2016})}\BibitemShut
  {NoStop}%
\bibitem [{\citenamefont {{Guillet}}\ \emph {et~al.}(2021)\citenamefont
  {{Guillet}}, \citenamefont {{Marty}}, \citenamefont {{Vergnaud}},
  \citenamefont {{Jamet}}, \citenamefont {{Zucchetti}}, \citenamefont
  {{Isella}}, \citenamefont {{Barbedienne}}, \citenamefont {{Jaffr{\`e}s}},
  \citenamefont {{Reyren}}, \citenamefont {{George}} \emph
  {et~al.}}]{guillet2021large}%
  \BibitemOpen
  \bibfield  {author} {\bibinfo {author} {\bibfnamefont {T.}~\bibnamefont
  {{Guillet}}}, \bibinfo {author} {\bibfnamefont {A.}~\bibnamefont {{Marty}}},
  \bibinfo {author} {\bibfnamefont {C.}~\bibnamefont {{Vergnaud}}}, \bibinfo
  {author} {\bibfnamefont {M.}~\bibnamefont {{Jamet}}}, \bibinfo {author}
  {\bibfnamefont {C.}~\bibnamefont {{Zucchetti}}}, \bibinfo {author}
  {\bibfnamefont {G.}~\bibnamefont {{Isella}}}, \bibinfo {author}
  {\bibfnamefont {Q.}~\bibnamefont {{Barbedienne}}}, \bibinfo {author}
  {\bibfnamefont {H.}~\bibnamefont {{Jaffr{\`e}s}}}, \bibinfo {author}
  {\bibfnamefont {N.}~\bibnamefont {{Reyren}}}, \bibinfo {author}
  {\bibfnamefont {J.~M.}\ \bibnamefont {{George}}},  \emph {et~al.},\ }\href
  {https://doi.org/10.1103/PhysRevB.103.064411} {\bibfield  {journal} {\bibinfo
   {journal} {\emph {Phys. Rev. B}},\ }\textbf {\bibinfo {volume} {103}},\
  \bibinfo {pages} {064411}\  (\bibinfo {year} {2021})}\BibitemShut {NoStop}%
\bibitem [{\citenamefont {\ifmmode~\check{Z}\else \v{Z}\fi{}elezn\'y}\ \emph
  {et~al.}(2021)\citenamefont {\ifmmode~\check{Z}\else \v{Z}\fi{}elezn\'y},
  \citenamefont {Fang}, \citenamefont {Olejn\'{\i}k}, \citenamefont {Patchett},
  \citenamefont {Gerhard}, \citenamefont {Gould}, \citenamefont {Molenkamp},
  \citenamefont {Gomez-Olivella}, \citenamefont {Zemen}, \citenamefont
  {Tich\'y} \emph {et~al.}}]{zelezny2021unidirectional}%
  \BibitemOpen
  \bibfield  {author} {\bibinfo {author} {\bibfnamefont {J.}~\bibnamefont
  {\ifmmode~\check{Z}\else \v{Z}\fi{}elezn\'y}}, \bibinfo {author}
  {\bibfnamefont {Z.}~\bibnamefont {Fang}}, \bibinfo {author} {\bibfnamefont
  {K.}~\bibnamefont {Olejn\'{\i}k}}, \bibinfo {author} {\bibfnamefont
  {J.}~\bibnamefont {Patchett}}, \bibinfo {author} {\bibfnamefont
  {F.}~\bibnamefont {Gerhard}}, \bibinfo {author} {\bibfnamefont
  {C.}~\bibnamefont {Gould}}, \bibinfo {author} {\bibfnamefont {L.~W.}\
  \bibnamefont {Molenkamp}}, \bibinfo {author} {\bibfnamefont {C.}~\bibnamefont
  {Gomez-Olivella}}, \bibinfo {author} {\bibfnamefont {J.}~\bibnamefont
  {Zemen}}, \bibinfo {author} {\bibfnamefont {T.}~\bibnamefont {Tich\'y}},
  \emph {et~al.},\ }\href
  {https://link.aps.org/doi/10.1103/PhysRevB.104.054429} {\bibfield  {journal}
  {\bibinfo  {journal} {\emph {Phys. Rev. B}},\ }\textbf {\bibinfo {volume}
  {104}},\ \bibinfo {pages} {054429}\  (\bibinfo {year} {2021})}\BibitemShut
  {NoStop}%
\bibitem [{\citenamefont {Zhou}\ \emph {et~al.}(2021)\citenamefont {Zhou},
  \citenamefont {Zeng}, \citenamefont {Jia}, \citenamefont {Chen},\ and\
  \citenamefont {Wu}}]{zhou2021sign}%
  \BibitemOpen
  \bibfield  {author} {\bibinfo {author} {\bibfnamefont {X.}~\bibnamefont
  {Zhou}}, \bibinfo {author} {\bibfnamefont {F.}~\bibnamefont {Zeng}}, \bibinfo
  {author} {\bibfnamefont {M.}~\bibnamefont {Jia}}, \bibinfo {author}
  {\bibfnamefont {H.}~\bibnamefont {Chen}}, \ and\ \bibinfo {author}
  {\bibfnamefont {Y.}~\bibnamefont {Wu}},\ }\href
  {https://link.aps.org/doi/10.1103/PhysRevB.104.184413} {\bibfield  {journal}
  {\bibinfo  {journal} {\emph {Phys. Rev. B}},\ }\textbf {\bibinfo {volume}
  {104}},\ \bibinfo {pages} {184413}\  (\bibinfo {year} {2021})}\BibitemShut
  {NoStop}%
\bibitem [{\citenamefont {Liu}\ \emph {et~al.}(2021)\citenamefont {Liu},
  \citenamefont {Wang}, \citenamefont {Luan}, \citenamefont {Zhou},
  \citenamefont {Xia}, \citenamefont {Yang}, \citenamefont {Tian},
  \citenamefont {Guo}, \citenamefont {Du},\ and\ \citenamefont
  {Wu}}]{liu2021magnonic}%
  \BibitemOpen
  \bibfield  {author} {\bibinfo {author} {\bibfnamefont {G.}~\bibnamefont
  {Liu}}, \bibinfo {author} {\bibfnamefont {X.-g.}\ \bibnamefont {Wang}},
  \bibinfo {author} {\bibfnamefont {Z.~Z.}\ \bibnamefont {Luan}}, \bibinfo
  {author} {\bibfnamefont {L.~F.}\ \bibnamefont {Zhou}}, \bibinfo {author}
  {\bibfnamefont {S.~Y.}\ \bibnamefont {Xia}}, \bibinfo {author} {\bibfnamefont
  {B.}~\bibnamefont {Yang}}, \bibinfo {author} {\bibfnamefont {Y.~Z.}\
  \bibnamefont {Tian}}, \bibinfo {author} {\bibfnamefont {G.-h.}\ \bibnamefont
  {Guo}}, \bibinfo {author} {\bibfnamefont {J.}~\bibnamefont {Du}}, \ and\
  \bibinfo {author} {\bibfnamefont {D.}~\bibnamefont {Wu}},\ }\href
  {https://link.aps.org/doi/10.1103/PhysRevLett.127.207206} {\bibfield
  {journal} {\bibinfo  {journal} {\emph {Phys. Rev. Lett.}},\ }\textbf
  {\bibinfo {volume} {127}},\ \bibinfo {pages} {207206}\  (\bibinfo {year}
  {2021})}\BibitemShut {NoStop}%
\bibitem [{\citenamefont {Nguyen}\ \emph {et~al.}(2021)\citenamefont {Nguyen},
  \citenamefont {Nguyen}, \citenamefont {Jeong}, \citenamefont {Park},
  \citenamefont {Jang}, \citenamefont {Lee}, \citenamefont {Lee}, \citenamefont
  {Park}, \citenamefont {Cho}, \citenamefont {Lee} \emph
  {et~al.}}]{nguyen2021unidirectional}%
  \BibitemOpen
  \bibfield  {author} {\bibinfo {author} {\bibfnamefont {T.~H.~T.}\
  \bibnamefont {Nguyen}}, \bibinfo {author} {\bibfnamefont {V.~Q.}\
  \bibnamefont {Nguyen}}, \bibinfo {author} {\bibfnamefont {S.}~\bibnamefont
  {Jeong}}, \bibinfo {author} {\bibfnamefont {E.}~\bibnamefont {Park}},
  \bibinfo {author} {\bibfnamefont {H.}~\bibnamefont {Jang}}, \bibinfo {author}
  {\bibfnamefont {N.~J.}\ \bibnamefont {Lee}}, \bibinfo {author} {\bibfnamefont
  {S.}~\bibnamefont {Lee}}, \bibinfo {author} {\bibfnamefont {B.-G.}\
  \bibnamefont {Park}}, \bibinfo {author} {\bibfnamefont {S.}~\bibnamefont
  {Cho}}, \bibinfo {author} {\bibfnamefont {H.-W.}\ \bibnamefont {Lee}},  \emph
  {et~al.},\ }\href {https://doi.org/10.1038/s42005-021-00743-9} {\bibfield
  {journal} {\bibinfo  {journal} {\emph {Commun. Phys.}},\ }\textbf {\bibinfo
  {volume} {4}},\ \bibinfo {pages} {247}\  (\bibinfo {year}
  {2021})}\BibitemShut {NoStop}%
\bibitem [{\citenamefont {Hasegawa}\ \emph {et~al.}(2021)\citenamefont
  {Hasegawa}, \citenamefont {Koyama},\ and\ \citenamefont
  {Chiba}}]{hasegawa2021enhanced}%
  \BibitemOpen
  \bibfield  {author} {\bibinfo {author} {\bibfnamefont {K.}~\bibnamefont
  {Hasegawa}}, \bibinfo {author} {\bibfnamefont {T.}~\bibnamefont {Koyama}}, \
  and\ \bibinfo {author} {\bibfnamefont {D.}~\bibnamefont {Chiba}},\ }\href
  {https://link.aps.org/doi/10.1103/PhysRevB.103.L020411} {\bibfield  {journal}
  {\bibinfo  {journal} {\emph {Phys. Rev. B}},\ }\textbf {\bibinfo {volume}
  {103}},\ \bibinfo {pages} {L020411}\  (\bibinfo {year} {2021})}\BibitemShut
  {NoStop}%
\bibitem [{\citenamefont {Mehraeen}\ and\ \citenamefont
  {Zhang}(2022)}]{mehraeen2022spin}%
  \BibitemOpen
  \bibfield  {author} {\bibinfo {author} {\bibfnamefont {M.}~\bibnamefont
  {Mehraeen}}\ and\ \bibinfo {author} {\bibfnamefont {S.~S.-L.}\ \bibnamefont
  {Zhang}},\ }\href {https://link.aps.org/doi/10.1103/PhysRevB.105.184423}
  {\bibfield  {journal} {\bibinfo  {journal} {\emph {Phys. Rev. B}},\ }\textbf
  {\bibinfo {volume} {105}},\ \bibinfo {pages} {184423}\  (\bibinfo {year}
  {2022})}\BibitemShut {NoStop}%
\bibitem [{\citenamefont {Shim}\ \emph {et~al.}(2022)\citenamefont {Shim},
  \citenamefont {Mehraeen}, \citenamefont {Sklenar}, \citenamefont {Oh},
  \citenamefont {Gibbons}, \citenamefont {Saglam}, \citenamefont {Hoffmann},
  \citenamefont {Zhang},\ and\ \citenamefont {Mason}}]{shim2022unidirectional}%
  \BibitemOpen
  \bibfield  {author} {\bibinfo {author} {\bibfnamefont {S.}~\bibnamefont
  {Shim}}, \bibinfo {author} {\bibfnamefont {M.}~\bibnamefont {Mehraeen}},
  \bibinfo {author} {\bibfnamefont {J.}~\bibnamefont {Sklenar}}, \bibinfo
  {author} {\bibfnamefont {J.}~\bibnamefont {Oh}}, \bibinfo {author}
  {\bibfnamefont {J.}~\bibnamefont {Gibbons}}, \bibinfo {author} {\bibfnamefont
  {H.}~\bibnamefont {Saglam}}, \bibinfo {author} {\bibfnamefont
  {A.}~\bibnamefont {Hoffmann}}, \bibinfo {author} {\bibfnamefont {S.~S.-L.}\
  \bibnamefont {Zhang}}, \ and\ \bibinfo {author} {\bibfnamefont
  {N.}~\bibnamefont {Mason}},\ }\href
  {https://link.aps.org/doi/10.1103/PhysRevX.12.021069} {\bibfield  {journal}
  {\bibinfo  {journal} {\emph {Phys. Rev. X}},\ }\textbf {\bibinfo {volume}
  {12}},\ \bibinfo {pages} {021069}\  (\bibinfo {year} {2022})}\BibitemShut
  {NoStop}%
\bibitem [{\citenamefont {Lou}\ \emph {et~al.}(2022)\citenamefont {Lou},
  \citenamefont {Zhao}, \citenamefont {Jiang},\ and\ \citenamefont
  {Bi}}]{lou2022large}%
  \BibitemOpen
  \bibfield  {author} {\bibinfo {author} {\bibfnamefont {K.}~\bibnamefont
  {Lou}}, \bibinfo {author} {\bibfnamefont {Q.}~\bibnamefont {Zhao}}, \bibinfo
  {author} {\bibfnamefont {B.}~\bibnamefont {Jiang}}, \ and\ \bibinfo {author}
  {\bibfnamefont {C.}~\bibnamefont {Bi}},\ }\href
  {https://link.aps.org/doi/10.1103/PhysRevApplied.17.064052} {\bibfield
  {journal} {\bibinfo  {journal} {\emph {Phys. Rev. Appl.}},\ }\textbf
  {\bibinfo {volume} {17}},\ \bibinfo {pages} {064052}\  (\bibinfo {year}
  {2022})}\BibitemShut {NoStop}%
\bibitem [{\citenamefont {Ding}\ \emph {et~al.}(2022)\citenamefont {Ding},
  \citenamefont {No\"el}, \citenamefont {Krishnaswamy},\ and\ \citenamefont
  {Gambardella}}]{ding2022unidirectional}%
  \BibitemOpen
  \bibfield  {author} {\bibinfo {author} {\bibfnamefont {S.}~\bibnamefont
  {Ding}}, \bibinfo {author} {\bibfnamefont {P.}~\bibnamefont {No\"el}},
  \bibinfo {author} {\bibfnamefont {G.~K.}\ \bibnamefont {Krishnaswamy}}, \
  and\ \bibinfo {author} {\bibfnamefont {P.}~\bibnamefont {Gambardella}},\
  }\href {https://link.aps.org/doi/10.1103/PhysRevResearch.4.L032041}
  {\bibfield  {journal} {\bibinfo  {journal} {\emph {Phys. Rev. Res.}},\
  }\textbf {\bibinfo {volume} {4}},\ \bibinfo {pages} {L032041}\  (\bibinfo
  {year} {2022})}\BibitemShut {NoStop}%
\bibitem [{\citenamefont {Fan}\ \emph {et~al.}(2022)\citenamefont {Fan},
  \citenamefont {Zhang}, \citenamefont {Han}, \citenamefont {Lv}, \citenamefont
  {Liu},\ and\ \citenamefont {Wang}}]{fan2022observation}%
  \BibitemOpen
  \bibfield  {author} {\bibinfo {author} {\bibfnamefont {Y.}~\bibnamefont
  {Fan}}, \bibinfo {author} {\bibfnamefont {P.}~\bibnamefont {Zhang}}, \bibinfo
  {author} {\bibfnamefont {J.}~\bibnamefont {Han}}, \bibinfo {author}
  {\bibfnamefont {Y.}~\bibnamefont {Lv}}, \bibinfo {author} {\bibfnamefont
  {L.}~\bibnamefont {Liu}}, \ and\ \bibinfo {author} {\bibfnamefont {J.-P.}\
  \bibnamefont {Wang}},\ }\href {https://doi.org/10.1002/aelm.202300232}
  {\bibfield  {journal} {\bibinfo  {journal} {\emph {Adv. Electron. Mater.}},\
  \bibinfo {pages} {2300232}}\  (\bibinfo {year} {2022})}\BibitemShut {NoStop}%
\bibitem [{\citenamefont {Cheng}\ \emph {et~al.}(2023)\citenamefont {Cheng},
  \citenamefont {Tang}, \citenamefont {Michel}, \citenamefont {Chong},
  \citenamefont {Yang}, \citenamefont {Cheng},\ and\ \citenamefont
  {Wang}}]{cheng2023unidirectional}%
  \BibitemOpen
  \bibfield  {author} {\bibinfo {author} {\bibfnamefont {Y.}~\bibnamefont
  {Cheng}}, \bibinfo {author} {\bibfnamefont {J.}~\bibnamefont {Tang}},
  \bibinfo {author} {\bibfnamefont {J.~J.}\ \bibnamefont {Michel}}, \bibinfo
  {author} {\bibfnamefont {S.~K.}\ \bibnamefont {Chong}}, \bibinfo {author}
  {\bibfnamefont {F.}~\bibnamefont {Yang}}, \bibinfo {author} {\bibfnamefont
  {R.}~\bibnamefont {Cheng}}, \ and\ \bibinfo {author} {\bibfnamefont {K.~L.}\
  \bibnamefont {Wang}},\ }\href
  {https://link.aps.org/doi/10.1103/PhysRevLett.130.086703} {\bibfield
  {journal} {\bibinfo  {journal} {\emph {Phys. Rev. Lett.}},\ }\textbf
  {\bibinfo {volume} {130}},\ \bibinfo {pages} {086703}\  (\bibinfo {year}
  {2023})}\BibitemShut {NoStop}%
\bibitem [{\citenamefont {Wang}\ \emph {et~al.}(2023)\citenamefont {Wang},
  \citenamefont {Cui}, \citenamefont {Xie}, \citenamefont {Zhang},
  \citenamefont {Tian}, \citenamefont {Bai}, \citenamefont {Huang},
  \citenamefont {Cao},\ and\ \citenamefont {Yan}}]{wang2023controllable}%
  \BibitemOpen
  \bibfield  {author} {\bibinfo {author} {\bibfnamefont {S.}~\bibnamefont
  {Wang}}, \bibinfo {author} {\bibfnamefont {X.}~\bibnamefont {Cui}}, \bibinfo
  {author} {\bibfnamefont {R.}~\bibnamefont {Xie}}, \bibinfo {author}
  {\bibfnamefont {C.}~\bibnamefont {Zhang}}, \bibinfo {author} {\bibfnamefont
  {Y.}~\bibnamefont {Tian}}, \bibinfo {author} {\bibfnamefont {L.}~\bibnamefont
  {Bai}}, \bibinfo {author} {\bibfnamefont {Q.}~\bibnamefont {Huang}}, \bibinfo
  {author} {\bibfnamefont {Q.}~\bibnamefont {Cao}}, \ and\ \bibinfo {author}
  {\bibfnamefont {S.}~\bibnamefont {Yan}},\ }\href
  {https://link.aps.org/doi/10.1103/PhysRevB.107.094410} {\bibfield  {journal}
  {\bibinfo  {journal} {\emph {Phys. Rev. B}},\ }\textbf {\bibinfo {volume}
  {107}},\ \bibinfo {pages} {094410}\  (\bibinfo {year} {2023})}\BibitemShut
  {NoStop}%
\bibitem [{\citenamefont {Zheng}\ \emph {et~al.}(2023)\citenamefont {Zheng},
  \citenamefont {Gu}, \citenamefont {Zhang}, \citenamefont {Zhang},
  \citenamefont {Zhao}, \citenamefont {Li}, \citenamefont {Ren}, \citenamefont
  {Jia}, \citenamefont {Xiao}, \citenamefont {Zhou} \emph
  {et~al.}}]{zheng2023coexistence}%
  \BibitemOpen
  \bibfield  {author} {\bibinfo {author} {\bibfnamefont {Z.}~\bibnamefont
  {Zheng}}, \bibinfo {author} {\bibfnamefont {Y.}~\bibnamefont {Gu}}, \bibinfo
  {author} {\bibfnamefont {Z.}~\bibnamefont {Zhang}}, \bibinfo {author}
  {\bibfnamefont {X.}~\bibnamefont {Zhang}}, \bibinfo {author} {\bibfnamefont
  {T.}~\bibnamefont {Zhao}}, \bibinfo {author} {\bibfnamefont {H.}~\bibnamefont
  {Li}}, \bibinfo {author} {\bibfnamefont {L.}~\bibnamefont {Ren}}, \bibinfo
  {author} {\bibfnamefont {L.}~\bibnamefont {Jia}}, \bibinfo {author}
  {\bibfnamefont {R.}~\bibnamefont {Xiao}}, \bibinfo {author} {\bibfnamefont
  {H.-A.}\ \bibnamefont {Zhou}},  \emph {et~al.},\ }\href
  {https://doi.org/10.1021/acs.nanolett.3c01082} {\bibfield  {journal}
  {\bibinfo  {journal} {\emph {Nano Lett.}}}\  (\bibinfo {year}
  {2023})}\BibitemShut {NoStop}%
\bibitem [{\citenamefont {Mehraeen}\ \emph {et~al.}(2023)\citenamefont
  {Mehraeen}, \citenamefont {Shen},\ and\ \citenamefont
  {Zhang}}]{mehraeen2023quantum}%
  \BibitemOpen
  \bibfield  {author} {\bibinfo {author} {\bibfnamefont {M.}~\bibnamefont
  {Mehraeen}}, \bibinfo {author} {\bibfnamefont {P.}~\bibnamefont {Shen}}, \
  and\ \bibinfo {author} {\bibfnamefont {S.~S.-L.}\ \bibnamefont {Zhang}},\
  }\href {https://link.aps.org/doi/10.1103/PhysRevB.108.014411} {\bibfield
  {journal} {\bibinfo  {journal} {\emph {Phys. Rev. B}},\ }\textbf {\bibinfo
  {volume} {108}},\ \bibinfo {pages} {014411}\  (\bibinfo {year}
  {2023})}\BibitemShut {NoStop}%
\bibitem [{Note1()}]{Note1}%
  \BibitemOpen
  \bibinfo {note} {It should be stressed that the SOC disorder is a required
  element, as without it, the in-plane magnetization--a key UMR ingredient--can
  be gauged out of the problem \cite {chiba2017magnetic,
  dyrdal2020spin}.}\BibitemShut {Stop}%
\bibitem [{\citenamefont {Sherman}(2003)}]{sherman2003minimum}%
  \BibitemOpen
  \bibfield  {author} {\bibinfo {author} {\bibfnamefont {E.~Y.}\ \bibnamefont
  {Sherman}},\ }\href {https://link.aps.org/doi/10.1103/PhysRevB.67.161303}
  {\bibfield  {journal} {\bibinfo  {journal} {\emph {Phys. Rev. B}},\ }\textbf
  {\bibinfo {volume} {67}},\ \bibinfo {pages} {161303}\  (\bibinfo {year}
  {2003})}\BibitemShut {NoStop}%
\bibitem [{\citenamefont {Golub}\ and\ \citenamefont
  {Ivchenko}(2004)}]{golub2004spin}%
  \BibitemOpen
  \bibfield  {author} {\bibinfo {author} {\bibfnamefont {L.~E.}\ \bibnamefont
  {Golub}}\ and\ \bibinfo {author} {\bibfnamefont {E.~L.}\ \bibnamefont
  {Ivchenko}},\ }\href {https://link.aps.org/doi/10.1103/PhysRevB.69.115333}
  {\bibfield  {journal} {\bibinfo  {journal} {\emph {Phys. Rev. B}},\ }\textbf
  {\bibinfo {volume} {69}},\ \bibinfo {pages} {115333}\  (\bibinfo {year}
  {2004})}\BibitemShut {NoStop}%
\bibitem [{\citenamefont {Str\"om}\ \emph {et~al.}(2010)\citenamefont
  {Str\"om}, \citenamefont {Johannesson},\ and\ \citenamefont
  {Japaridze}}]{strom2010edge}%
  \BibitemOpen
  \bibfield  {author} {\bibinfo {author} {\bibfnamefont {A.}~\bibnamefont
  {Str\"om}}, \bibinfo {author} {\bibfnamefont {H.}~\bibnamefont
  {Johannesson}}, \ and\ \bibinfo {author} {\bibfnamefont {G.~I.}\ \bibnamefont
  {Japaridze}},\ }\href
  {https://link.aps.org/doi/10.1103/PhysRevLett.104.256804} {\bibfield
  {journal} {\bibinfo  {journal} {\emph {Phys. Rev. Lett.}},\ }\textbf
  {\bibinfo {volume} {104}},\ \bibinfo {pages} {256804}\  (\bibinfo {year}
  {2010})}\BibitemShut {NoStop}%
\bibitem [{\citenamefont {Kimme}\ \emph {et~al.}(2016)\citenamefont {Kimme},
  \citenamefont {Rosenow},\ and\ \citenamefont
  {Brataas}}]{kimme2016backscattering}%
  \BibitemOpen
  \bibfield  {author} {\bibinfo {author} {\bibfnamefont {L.}~\bibnamefont
  {Kimme}}, \bibinfo {author} {\bibfnamefont {B.}~\bibnamefont {Rosenow}}, \
  and\ \bibinfo {author} {\bibfnamefont {A.}~\bibnamefont {Brataas}},\ }\href
  {https://link.aps.org/doi/10.1103/PhysRevB.93.081301} {\bibfield  {journal}
  {\bibinfo  {journal} {\emph {Phys. Rev. B}},\ }\textbf {\bibinfo {volume}
  {93}},\ \bibinfo {pages} {081301}\  (\bibinfo {year} {2016})}\BibitemShut
  {NoStop}%
\bibitem [{\citenamefont {Dyrda\l{}}\ \emph {et~al.}(2020)\citenamefont
  {Dyrda\l{}}, \citenamefont {Barna\ifmmode~\acute{s}\else \'{s}\fi{}},\ and\
  \citenamefont {Fert}}]{dyrdal2020spin}%
  \BibitemOpen
  \bibfield  {author} {\bibinfo {author} {\bibfnamefont {A.}~\bibnamefont
  {Dyrda\l{}}}, \bibinfo {author} {\bibfnamefont {J.}~\bibnamefont
  {Barna\ifmmode~\acute{s}\else \'{s}\fi{}}}, \ and\ \bibinfo {author}
  {\bibfnamefont {A.}~\bibnamefont {Fert}},\ }\href
  {https://link.aps.org/doi/10.1103/PhysRevLett.124.046802} {\bibfield
  {journal} {\bibinfo  {journal} {\emph {Phys. Rev. Lett.}},\ }\textbf
  {\bibinfo {volume} {124}},\ \bibinfo {pages} {046802}\  (\bibinfo {year}
  {2020})}\BibitemShut {NoStop}%
\bibitem [{\citenamefont {Kubo}(1957)}]{kubo1957statistical}%
  \BibitemOpen
  \bibfield  {author} {\bibinfo {author} {\bibfnamefont {R.}~\bibnamefont
  {Kubo}},\ }\href {https://doi.org/10.1143/JPSJ.12.570} {\bibfield  {journal}
  {\bibinfo  {journal} {\emph {J. Phys. Soc. Jpn.}},\ }\textbf {\bibinfo
  {volume} {12}},\ \bibinfo {pages} {570}\  (\bibinfo {year}
  {1957})}\BibitemShut {NoStop}%
\bibitem [{\citenamefont {Parker}\ \emph {et~al.}(2019)\citenamefont {Parker},
  \citenamefont {Morimoto}, \citenamefont {Orenstein},\ and\ \citenamefont
  {Moore}}]{parker2019diagrammatic}%
  \BibitemOpen
  \bibfield  {author} {\bibinfo {author} {\bibfnamefont {D.~E.}\ \bibnamefont
  {Parker}}, \bibinfo {author} {\bibfnamefont {T.}~\bibnamefont {Morimoto}},
  \bibinfo {author} {\bibfnamefont {J.}~\bibnamefont {Orenstein}}, \ and\
  \bibinfo {author} {\bibfnamefont {J.~E.}\ \bibnamefont {Moore}},\ }\href
  {https://link.aps.org/doi/10.1103/PhysRevB.99.045121} {\bibfield  {journal}
  {\bibinfo  {journal} {\emph {Phys. Rev. B}},\ }\textbf {\bibinfo {volume}
  {99}},\ \bibinfo {pages} {045121}\  (\bibinfo {year} {2019})}\BibitemShut
  {NoStop}%
\bibitem [{\citenamefont {Du}\ \emph {et~al.}(2021)\citenamefont {Du},
  \citenamefont {Wang}, \citenamefont {Sun}, \citenamefont {Lu},\ and\
  \citenamefont {Xie}}]{du2021quantum}%
  \BibitemOpen
  \bibfield  {author} {\bibinfo {author} {\bibfnamefont {Z.}~\bibnamefont
  {Du}}, \bibinfo {author} {\bibfnamefont {C.}~\bibnamefont {Wang}}, \bibinfo
  {author} {\bibfnamefont {H.-P.}\ \bibnamefont {Sun}}, \bibinfo {author}
  {\bibfnamefont {H.-Z.}\ \bibnamefont {Lu}}, \ and\ \bibinfo {author}
  {\bibfnamefont {X.}~\bibnamefont {Xie}},\ }\href
  {https://doi.org/10.1038/s41467-021-25273-4} {\bibfield  {journal} {\bibinfo
  {journal} {\emph {Nat. Commun.}},\ }\textbf {\bibinfo {volume} {12}},\
  \bibinfo {pages} {1}\  (\bibinfo {year} {2021})}\BibitemShut {NoStop}%
\bibitem [{\citenamefont {Rostami}\ \emph {et~al.}(2021)\citenamefont
  {Rostami}, \citenamefont {Katsnelson}, \citenamefont {Vignale},\ and\
  \citenamefont {Polini}}]{rostami2021gauge}%
  \BibitemOpen
  \bibfield  {author} {\bibinfo {author} {\bibfnamefont {H.}~\bibnamefont
  {Rostami}}, \bibinfo {author} {\bibfnamefont {M.~I.}\ \bibnamefont
  {Katsnelson}}, \bibinfo {author} {\bibfnamefont {G.}~\bibnamefont {Vignale}},
  \ and\ \bibinfo {author} {\bibfnamefont {M.}~\bibnamefont {Polini}},\ }\href
  {https://doi.org/10.1016/j.aop.2021.168523} {\bibfield  {journal} {\bibinfo
  {journal} {\emph {Ann. Phys.}},\ }\textbf {\bibinfo {volume} {431}},\
  \bibinfo {pages} {168523}\  (\bibinfo {year} {2021})}\BibitemShut {NoStop}%
\bibitem [{\citenamefont {Qi}\ and\ \citenamefont
  {Zhang}(2011)}]{qi2011topological}%
  \BibitemOpen
  \bibfield  {author} {\bibinfo {author} {\bibfnamefont {X.-L.}\ \bibnamefont
  {Qi}}\ and\ \bibinfo {author} {\bibfnamefont {S.-C.}\ \bibnamefont {Zhang}},\
  }\href {https://link.aps.org/doi/10.1103/RevModPhys.83.1057} {\bibfield
  {journal} {\bibinfo  {journal} {\emph {Rev. Mod. Phys.}},\ }\textbf {\bibinfo
  {volume} {83}},\ \bibinfo {pages} {1057}\  (\bibinfo {year}
  {2011})}\BibitemShut {NoStop}%
\bibitem [{Note2()}]{Note2}%
  \BibitemOpen
  \bibinfo {note} {It is worth noting that these angular dependence trends of
  the NPHE for both out-of-plane and in-plane sweeps of the magnetization
  resemble the behavior predicted for the planar Hall effect in the linear
  response regime \cite {chiba2017magnetic}.}\BibitemShut {Stop}%
\bibitem [{\citenamefont {He}\ \emph {et~al.}(2018)\citenamefont {He},
  \citenamefont {Zhang}, \citenamefont {Zhu}, \citenamefont {Liu},
  \citenamefont {Wang}, \citenamefont {Yu}, \citenamefont {Vignale},\ and\
  \citenamefont {Yang}}]{he2018bilinear}%
  \BibitemOpen
  \bibfield  {author} {\bibinfo {author} {\bibfnamefont {P.}~\bibnamefont
  {He}}, \bibinfo {author} {\bibfnamefont {S.~S.-L.}\ \bibnamefont {Zhang}},
  \bibinfo {author} {\bibfnamefont {D.}~\bibnamefont {Zhu}}, \bibinfo {author}
  {\bibfnamefont {Y.}~\bibnamefont {Liu}}, \bibinfo {author} {\bibfnamefont
  {Y.}~\bibnamefont {Wang}}, \bibinfo {author} {\bibfnamefont {J.}~\bibnamefont
  {Yu}}, \bibinfo {author} {\bibfnamefont {G.}~\bibnamefont {Vignale}}, \ and\
  \bibinfo {author} {\bibfnamefont {H.}~\bibnamefont {Yang}},\ }\href
  {https://doi.org/10.1038/s41567-017-0039-y} {\bibfield  {journal} {\bibinfo
  {journal} {\emph {Nat. Phys.}},\ }\textbf {\bibinfo {volume} {14}},\ \bibinfo
  {pages} {495}\  (\bibinfo {year} {2018})}\BibitemShut {NoStop}%
\bibitem [{\citenamefont {He}\ \emph {et~al.}(2019)\citenamefont {He},
  \citenamefont {Zhang}, \citenamefont {Zhu}, \citenamefont {Shi},
  \citenamefont {Heinonen}, \citenamefont {Vignale},\ and\ \citenamefont
  {Yang}}]{he2019nonlinear}%
  \BibitemOpen
  \bibfield  {author} {\bibinfo {author} {\bibfnamefont {P.}~\bibnamefont
  {He}}, \bibinfo {author} {\bibfnamefont {S.~S.-L.}\ \bibnamefont {Zhang}},
  \bibinfo {author} {\bibfnamefont {D.}~\bibnamefont {Zhu}}, \bibinfo {author}
  {\bibfnamefont {S.}~\bibnamefont {Shi}}, \bibinfo {author} {\bibfnamefont
  {O.~G.}\ \bibnamefont {Heinonen}}, \bibinfo {author} {\bibfnamefont
  {G.}~\bibnamefont {Vignale}}, \ and\ \bibinfo {author} {\bibfnamefont
  {H.}~\bibnamefont {Yang}},\ }\href
  {https://link.aps.org/doi/10.1103/PhysRevLett.123.016801} {\bibfield
  {journal} {\bibinfo  {journal} {\emph {Phys. Rev. Lett.}},\ }\textbf
  {\bibinfo {volume} {123}},\ \bibinfo {pages} {016801}\  (\bibinfo {year}
  {2019})}\BibitemShut {NoStop}%
\bibitem [{\citenamefont {Langenfeld}\ \emph {et~al.}(2016)\citenamefont
  {Langenfeld}, \citenamefont {Tshitoyan}, \citenamefont {Fang}, \citenamefont
  {Wells}, \citenamefont {Moore},\ and\ \citenamefont
  {Ferguson}}]{langenfeld2016exchange}%
  \BibitemOpen
  \bibfield  {author} {\bibinfo {author} {\bibfnamefont {S.}~\bibnamefont
  {Langenfeld}}, \bibinfo {author} {\bibfnamefont {V.}~\bibnamefont
  {Tshitoyan}}, \bibinfo {author} {\bibfnamefont {Z.}~\bibnamefont {Fang}},
  \bibinfo {author} {\bibfnamefont {A.}~\bibnamefont {Wells}}, \bibinfo
  {author} {\bibfnamefont {T.}~\bibnamefont {Moore}}, \ and\ \bibinfo {author}
  {\bibfnamefont {A.}~\bibnamefont {Ferguson}},\ }\href
  {https://doi.org/10.1063/1.4948921} {\bibfield  {journal} {\bibinfo
  {journal} {\emph {Appl. Phys. Lett.}},\ }\textbf {\bibinfo {volume} {108}}\
  (\bibinfo {year} {2016})}\BibitemShut {NoStop}%
\bibitem [{\citenamefont {Sterk}\ \emph {et~al.}(2019)\citenamefont {Sterk},
  \citenamefont {Peerlings},\ and\ \citenamefont {Duine}}]{sterk2019magnon}%
  \BibitemOpen
  \bibfield  {author} {\bibinfo {author} {\bibfnamefont {W.~P.}\ \bibnamefont
  {Sterk}}, \bibinfo {author} {\bibfnamefont {D.}~\bibnamefont {Peerlings}}, \
  and\ \bibinfo {author} {\bibfnamefont {R.~A.}\ \bibnamefont {Duine}},\ }\href
  {https://link.aps.org/doi/10.1103/PhysRevB.99.064438} {\bibfield  {journal}
  {\bibinfo  {journal} {\emph {Phys. Rev. B}},\ }\textbf {\bibinfo {volume}
  {99}},\ \bibinfo {pages} {064438}\  (\bibinfo {year} {2019})}\BibitemShut
  {NoStop}%
\bibitem [{\citenamefont {Sodemann}\ and\ \citenamefont
  {Fu}(2015)}]{sodemann2015quantum}%
  \BibitemOpen
  \bibfield  {author} {\bibinfo {author} {\bibfnamefont {I.}~\bibnamefont
  {Sodemann}}\ and\ \bibinfo {author} {\bibfnamefont {L.}~\bibnamefont {Fu}},\
  }\href {https://link.aps.org/doi/10.1103/PhysRevLett.115.216806} {\bibfield
  {journal} {\bibinfo  {journal} {\emph {Phys. Rev. Lett.}},\ }\textbf
  {\bibinfo {volume} {115}},\ \bibinfo {pages} {216806}\  (\bibinfo {year}
  {2015})}\BibitemShut {NoStop}%
\bibitem [{\citenamefont {Zhang}\ \emph {et~al.}(2019)\citenamefont {Zhang},
  \citenamefont {Shi}, \citenamefont {Zhu}, \citenamefont {Xing}, \citenamefont
  {Zhang},\ and\ \citenamefont {Wang}}]{zhang2019topological}%
  \BibitemOpen
  \bibfield  {author} {\bibinfo {author} {\bibfnamefont {D.}~\bibnamefont
  {Zhang}}, \bibinfo {author} {\bibfnamefont {M.}~\bibnamefont {Shi}}, \bibinfo
  {author} {\bibfnamefont {T.}~\bibnamefont {Zhu}}, \bibinfo {author}
  {\bibfnamefont {D.}~\bibnamefont {Xing}}, \bibinfo {author} {\bibfnamefont
  {H.}~\bibnamefont {Zhang}}, \ and\ \bibinfo {author} {\bibfnamefont
  {J.}~\bibnamefont {Wang}},\ }\href
  {https://link.aps.org/doi/10.1103/PhysRevLett.122.206401} {\bibfield
  {journal} {\bibinfo  {journal} {\emph {Phys. Rev. Lett.}},\ }\textbf
  {\bibinfo {volume} {122}},\ \bibinfo {pages} {206401}\  (\bibinfo {year}
  {2019})}\BibitemShut {NoStop}%
\bibitem [{\citenamefont {Wang}\ \emph {et~al.}(2020)\citenamefont {Wang},
  \citenamefont {Wang}, \citenamefont {Yang}, \citenamefont {Shi},
  \citenamefont {Ruan}, \citenamefont {Xing}, \citenamefont {Wang},\ and\
  \citenamefont {Zhang}}]{wang2020dynamical}%
  \BibitemOpen
  \bibfield  {author} {\bibinfo {author} {\bibfnamefont {H.}~\bibnamefont
  {Wang}}, \bibinfo {author} {\bibfnamefont {D.}~\bibnamefont {Wang}}, \bibinfo
  {author} {\bibfnamefont {Z.}~\bibnamefont {Yang}}, \bibinfo {author}
  {\bibfnamefont {M.}~\bibnamefont {Shi}}, \bibinfo {author} {\bibfnamefont
  {J.}~\bibnamefont {Ruan}}, \bibinfo {author} {\bibfnamefont {D.}~\bibnamefont
  {Xing}}, \bibinfo {author} {\bibfnamefont {J.}~\bibnamefont {Wang}}, \ and\
  \bibinfo {author} {\bibfnamefont {H.}~\bibnamefont {Zhang}},\ }\href
  {https://link.aps.org/doi/10.1103/PhysRevB.101.081109} {\bibfield  {journal}
  {\bibinfo  {journal} {\emph {Phys. Rev. B}},\ }\textbf {\bibinfo {volume}
  {101}},\ \bibinfo {pages} {081109}\  (\bibinfo {year} {2020})}\BibitemShut
  {NoStop}%
\bibitem [{\citenamefont {Liu}\ \emph {et~al.}(2009)\citenamefont {Liu},
  \citenamefont {Liu}, \citenamefont {Xu}, \citenamefont {Qi},\ and\
  \citenamefont {Zhang}}]{liu2009magnetic}%
  \BibitemOpen
  \bibfield  {author} {\bibinfo {author} {\bibfnamefont {Q.}~\bibnamefont
  {Liu}}, \bibinfo {author} {\bibfnamefont {C.-X.}\ \bibnamefont {Liu}},
  \bibinfo {author} {\bibfnamefont {C.}~\bibnamefont {Xu}}, \bibinfo {author}
  {\bibfnamefont {X.-L.}\ \bibnamefont {Qi}}, \ and\ \bibinfo {author}
  {\bibfnamefont {S.-C.}\ \bibnamefont {Zhang}},\ }\href
  {https://link.aps.org/doi/10.1103/PhysRevLett.102.156603} {\bibfield
  {journal} {\bibinfo  {journal} {\emph {Phys. Rev. Lett.}},\ }\textbf
  {\bibinfo {volume} {102}},\ \bibinfo {pages} {156603}\  (\bibinfo {year}
  {2009})}\BibitemShut {NoStop}%
\bibitem [{\citenamefont {Abanin}\ and\ \citenamefont
  {Pesin}(2011)}]{abanin2011ordering}%
  \BibitemOpen
  \bibfield  {author} {\bibinfo {author} {\bibfnamefont {D.~A.}\ \bibnamefont
  {Abanin}}\ and\ \bibinfo {author} {\bibfnamefont {D.~A.}\ \bibnamefont
  {Pesin}},\ }\href {https://link.aps.org/doi/10.1103/PhysRevLett.106.136802}
  {\bibfield  {journal} {\bibinfo  {journal} {\emph {Phys. Rev. Lett.}},\
  }\textbf {\bibinfo {volume} {106}},\ \bibinfo {pages} {136802}\  (\bibinfo
  {year} {2011})}\BibitemShut {NoStop}%
\bibitem [{\citenamefont {Nomura}\ and\ \citenamefont
  {Nagaosa}(2011)}]{nomura2011surface}%
  \BibitemOpen
  \bibfield  {author} {\bibinfo {author} {\bibfnamefont {K.}~\bibnamefont
  {Nomura}}\ and\ \bibinfo {author} {\bibfnamefont {N.}~\bibnamefont
  {Nagaosa}},\ }\href {https://link.aps.org/doi/10.1103/PhysRevLett.106.166802}
  {\bibfield  {journal} {\bibinfo  {journal} {\emph {Phys. Rev. Lett.}},\
  }\textbf {\bibinfo {volume} {106}},\ \bibinfo {pages} {166802}\  (\bibinfo
  {year} {2011})}\BibitemShut {NoStop}%
\end{thebibliography}%

\end{document}